\documentclass{aa}
\usepackage{epstopdf}
\usepackage[varg]{txfonts}
\usepackage{natbib}
\bibpunct{(}{)}{;}{a}{}{,} 
\usepackage[pdfborderstyle={/S/B}]{hyperref}
\usepackage{longtable}

\begin{document}

\title{The sharpness of gamma-ray burst prompt emission spectra}

\author{Hoi-Fung Yu\inst{\ref{inst1},\ref{inst2}} \and Hendrik J. van Eerten\inst{\ref{inst1}}\thanks{Fellow of the Alexander v. Humboldt Foundation} \and Jochen Greiner\inst{\ref{inst1},\ref{inst2}} \and Re'em Sari\inst{\ref{inst3}} \and P. Narayana Bhat\inst{\ref{inst4}} \and Andreas von Kienlin\inst{\ref{inst1}} \and William S. Paciesas\inst{\ref{inst5}} \and Robert D. Preece\inst{\ref{inst6}}}

\institute{Max-Planck-Institut f{\"u}r extraterrestrische Physik, Giessenbachstra{\ss}e 1, 85748 Garching, Germany\\ \email{sptfung@mpe.mpg.de}\label{inst1}
\and Excellence Cluster Universe, Technische Universit{\"a}t M{\"u}nchen, Boltzmannstra{\ss}e 2, 85748 Garching, Germany\label{inst2}
\and The Hebrew University of Jerusalem, Jerusalem, Israel\label{inst3}
\and Center for Space Plasma and Aeronomic Research, University of Alabama in Huntsville, Huntsville, AL 35805, USA\label{inst4}
\and Universities Space Research Association, Huntsville, AL 35805, USA\label{inst5}
\and Space Science Department, University of Alabama in Huntsville, Huntsville, AL 35809, USA\label{inst6}
}

\abstract
{We study the sharpness of the time-resolved prompt emission spectra of gamma-ray bursts (GRBs) observed by the Gamma-ray Burst Monitor (GBM) on board the \textit{Fermi} Gamma-ray Space Telescope.}
{We aim to obtain a measure of the curvature of time-resolved spectra that can be compared directly to theory. This tests the ability of models such as synchrotron emission to explain the peaks or breaks of GBM prompt emission spectra.}
{We take the burst sample from the official \textit{Fermi} GBM GRB time-resolved spectral catalog. We re-fit all spectra with a measured peak or break energy in the catalog best-fit models in various energy ranges, which cover the curvature around the spectral peak or break, resulting in a total of 1,113 spectra being analyzed. We compute the sharpness angles under the peak or break of the triangle constructed under the model fit curves and compare them to the values obtained from various representative emission models: blackbody, single-electron synchrotron, synchrotron emission from a Maxwellian or power-law electron distribution.}
{We find that 35\% of the time-resolved spectra are inconsistent with the single-electron synchrotron function, and 91\% are inconsistent with the Maxwellian synchrotron function. The single temperature, single emission time, and location blackbody function is found to be sharper than all the spectra. No general evolutionary trend of the sharpness angle is observed, neither per burst nor for the whole population. It is found that the limiting case, a single temperature Maxwellian synchrotron function, can only contribute up to $58_{-18}^{+23}$\% of the peak flux.}
{Our results show that even the sharpest but non-realistic case, the single-electron synchrotron function, cannot explain a large fraction of the observed GRB prompt spectra. Because any combination of physically possible synchrotron spectra added together will always further broaden the spectrum, emission mechanisms other than optically thin synchrotron radiation are likely required in a full explanation of the spectral peaks or breaks of the GRB prompt emission phase.}


\titlerunning{The sharpness of GRB prompt emission spectra}
\maketitle

\section{Introduction}
\label{sect:intro}

The prompt emission of gamma-ray bursts (GRBs) is one of the most puzzling observed astronomical phenomena. Since the discovery of GRBs in 1967 \citep{Klebesadel73a}, many emission models have been proposed in order to explain the prompt phase of gamma-ray emission. This phase consists of gamma rays mainly within tens to a few hundred keV, in some cases as high as a few thousand keV, lasting from a few milliseconds to hundreds of seconds.

Gamma-ray bursts are distributed isotropically \citep{Briggs96a,Hakkila94a,Tegmark96a} and cosmologically \citep{Metzger97a,Waxman97a} over the sky. Despite the last 45 years of research efforts, the dominant emission mechanism of these cosmological sources is still controversial. Synchrotron radiation from a simple electron population is one of the simplest physical phenomena that may be able to produce the observed spectral slopes of the Band function \citep{Band93a} that is commonly used to describe the photon spectra of GRB prompt emission. The Band function is an empirical mathematical function consisting of two segments of power laws, described by the low- and high-energy photon indices $\alpha$ and $\beta$, connected at the peak energy parameterized as $E_\text{p}$. This peak energy has been observed typically at hundreds of keV \citep{Kaneko06a,Nava11a,Goldstein12a,Goldstein13a,Gruber14a}. In what is known as the fireball model \citep{Goodman86a,Meszaros93a,Meszaros93b,Rees92a,Rees94a,Tavani96a,Piran99a}, there are ejected shells with different bulk Lorentz factors. When the faster shells catch up with the slower shells, internal shock waves will be produced. The electrons in the shocked region of the shells are accelerated and their energy is radiated via synchrotron emission in the local magnetic field.

The Band function's two power-law indices are usually compared to the slopes of various radiation models, leading to the discovery of the so-called line-of-death problem \citep{Katz94a,Crider98a,Preece98a,Preece02a,Tavani95a} for the synchrotron theory. When a power-law distribution of electron energies is combined with synchrotron radiation theory, the low-energy power-law photon index is $-2/3$. The fact that a fraction of observed $\alpha > -2/3$ indicates that, at least in some cases, the synchrotron explanation of GRB prompt spectra can be problematic, because the observed spectra rise faster. The observed violations of the line-of-death are typically around 30\%. Recently, \citet{Burgess15a} used the Band function to fit a large number of simulated slow-cooling synchrotron spectra (with spectral peak determined by injection energy of the electrons rather than their energy losses, as in the fast-cooling case), concluding that, in practice, the line-of-death may be steeper, $\alpha > -0.8$, than the value of $\alpha > -2/3$. Moreover, they found that the Band function cannot recover the simulated synchrotron peaks and power-law indices. These findings all question the validity of using the synchrotron theories to explain the Band parameters.

Instead of fitting the empirical Band function to the spectra, \citet{Burgess14a} used a synchrotron function in the fitting process, combining the slow-cooling scenario with thermal emission. \citet{Yu15a} used a double broken power law to fit eight bright GRBs, in which they found that the line-of-death problem could be alleviated in a moderately fast-cooling scenario, in which the fast- and slow-cooling electrons are mixed together, usually with a blackbody component at tens of keV or in a varying magnetic field. However, no single synchrotron model could completely explain all the spectral properties of GRB prompt emission.

In this work we study the sharpness of the synchrotron emission spectrum in comparison to time-resolved spectra of GRBs, a question recently raised by \citet{Beloborodov13a,Vurm15a}. Our approach focuses on the curvature of the region capturing the peak or break energy in the GRB prompt spectra by re-fitting all the spectra in an energy domain depending on this peak or break, using the burst sample from the \textit{Fermi} GBM time-resolved spectral catalog (Yu et al. in prep.). By comparing the spectral sharpness of the observed spectra to various physical emission models, we are able to directly determine whether a model is capable of accounting for the peak emission of the observed spectra. By concentrating on the spectral peak or break, we avoid potential issues with interpretation of the asymptotes of the fit functions, which might lie outside the observable domain or be contaminated by additional radiative processes or instrumental effects.

Thanks to the high-quality gamma-ray data obtained by \textit{Fermi} GBM \citep{Meegan09a}, which provides wide energy coverage and fine temporal and spectral resolutions, this is the first time that we can directly compare the curvature of a large number of time-resolved spectra to that of physical models, so that statistically significant conclusions about the prompt emission mechanism can be drawn.

This paper is structured as follows. In Section~\ref{sect:anal}, we describe our analysis method. The results are presented in Section~\ref{sect:resu}. In Section~\ref{sect:consistency}, we check the consistency of our analysis. In Section~\ref{sect:disc}, we discuss the theoretical implications. The summary and conclusions are given in Section~\ref{sect:conc}. Unless otherwise stated, all errors reported in this paper are given at the $1\sigma$ confidence level.

We note that a recent independent study of the peak-flux GRB spectra \citep{Axelsson15a} shows that a synchrotron function could be too wide for the observed Band shape. They measured the full-width-half-maximum in the $\nu F_\nu$ spectra obtained from the 4-years \textit{Fermi} GBM GRB time-integrated spectral catalog \citep{Gruber14a,vonKienlin14a} and the BATSE 5B GRB spectral catalog \citep{Goldstein13a}.

\section{Data analysis}
\label{sect:anal}

\subsection{The data and the method}

The \textit{Fermi} GBM consists of 12 thallium activated sodium iodide (NaI(Tl)) detectors, which cover 8~keV - 1~MeV, and 2 bismuth germanate (BGO) detectors, which cover 200~keV - 40~MeV. The combined energy range of the two kinds of detectors is ideal for the study of GRB prompt emission spectra because the typical spectrum peaks at a few hundred keV.

In order to account for the change in orientation of the source with respect to the detectors, due to the slew of the spacecraft, RSP2 files are used in the fitting process, which contain the detector response matrices for every two degrees on the sky. For each burst a low-order polynomial (order 2 - 4) is fit to every energy channel, according to a user-defined background interval before and after the prompt emission phase, and interpolated across the emission interval. All spectra are re-fit with the GBM official spectral analysis software RMFIT\footnote{The public version of the RMFIT software is available at \url{http://fermi.gsfc.nasa.gov/ssc/data/analysis/rmfit/}} v4.4.2BA and the GBM response matrices v2.0.

Our sample is taken from the official \textit{Fermi} GBM GRB time-resolved spectral catalog (Yu et al. in prep.) which consists of the brightest bursts observed by GBM before 21 August 2012. All the bursts in our sample are long bursts, i.e., with $T_{90} > 2$~s \citep{Kouveliotou93a}. They were selected according to 3 criteria: (1) the total fluence in 10~keV - 1~MeV, $f>4.0\times10^{-5}$~erg~cm$^{-2}$; (2) the peak flux in 10~keV - 1~MeV, $F_\text{p}>20$~ph~s$^{-1}$~cm$^{-2}$ in either 64, 256, or 1,024~ms binning timescales; and (3) the burst has 5 or more time bins when binned with signal-to-noise ratio $S/N=30$. This results in 81 bursts and 1,802 spectra in total, of which 311 do not satisfy the catalog's standard error criteria \citep[for the details on the error criteria, see][]{Gruber14a}. We exclude from further analysis these 311 spectra, and concentrate on the remaining 1,491 spectra.

Only the spectra best fit by the Comptonized model (COMP), the Band function (BAND), and the smoothly broken power law (SBPL) are included in the analysis. The functional forms of these models are given in Appendix~\ref{app:fitfunc} for completeness. This is because we are interested in comparing the sharpness around the peak or break energies of theoretical models to the observed spectra. Thus, the 194 spectra best fit by a simple power law or the power law plus blackbody, are excluded. All the 1,297 spectra best fit by either COMP, BAND, or SBPL have convex shape (i.e., $\alpha > -2$ for COMP and $\alpha > \beta$ for BAND and SBPL).

The best-fit model parameters for the 1,297 spectra are obtained from the catalog. Using the catalog values of $E_\text{p}$, every spectrum is re-fit (using the same best fit function as in the catalog) in a narrower energy domain that covers $E_\text{p}$ (or the break energy $E_\text{b}$ if there is no peak in the spectrum). We refer this energy domain as the "data domain", which contains the "triangle domain" (described below). We find that in the 1,297 spectra, 34 of them have no converged re-fit and are therefore excluded. In the remaining 1,263 spectra, 150 of them have large error bars \citep[according to the criteria from][]{Gruber14a} and thus are further excluded. In total, 1,113 spectra are used in this work, of which 942 are best fit by COMP, 99 by BAND, and 72 by SBPL.

Motivated by the necessity to model the spectral curvature around $E_\text{p}$ and exclude any possible curvature contribution from the low- or high-energy tail, we test the data domain on a few of the brightest bursts (details are discussed in Sect.~\ref{sect:datadomain}). As a result, $(E_\text{left}, E_\text{right}) = (0.1 E_\text{p}, 3.0 E_\text{p})$ is adopted for a good balance between statistics and optimal description of the spectral sharpness. Assuming a typical spectrum with $E_\text{p}\sim300$~keV, it means that we are covering the range from $\sim30$~keV to $\sim900$~keV.

\begin{figure}
\resizebox{\hsize}{!}
{\includegraphics{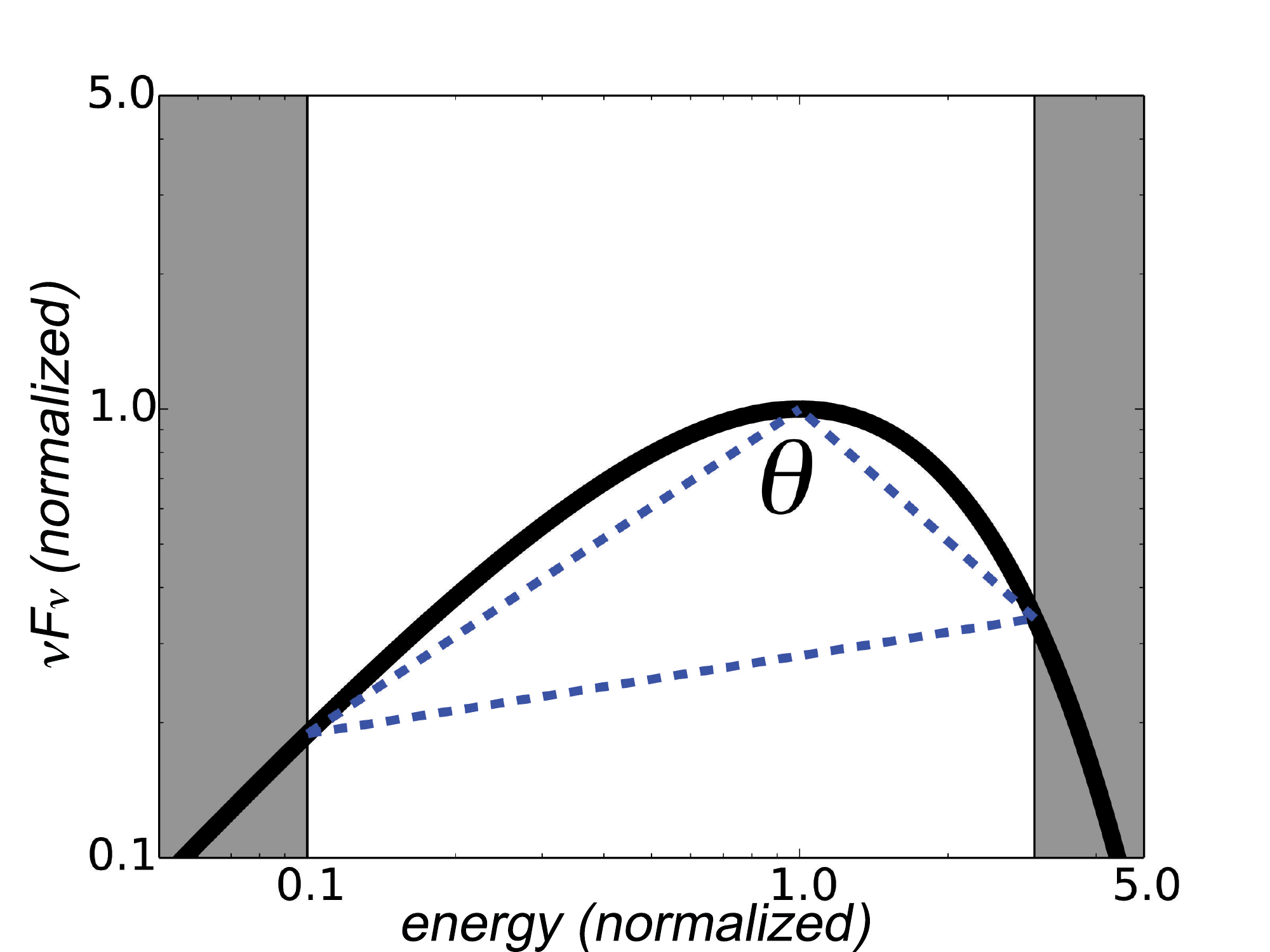}}
\caption{Illustration of how the triangle is constructed and the sharpness angle $\theta$ is defined. The shaded regions indicate the data domain (see Sect.~\ref{sect:datadomain}). The triangle is constructed within the triangle domain (see Sect.~\ref{sect:angleodmain}), under the best-fit model curve (black). The vertical and horizontal axis are plotted in logarithmic scale in units of normalized $\nu F_\nu$ flux and photon energy, respectively.}
\label{fig:triangle}
\end{figure}

The peak energy $E_\text{p}$ and the peak flux $\nu F_\nu(E_\text{p})$ are used to normalize the model curve such that the peak coincides with $(x,y) = (E/E_\text{p},\nu F_\nu(E)/\nu F_\nu(E_\text{p})) = (1,1)$. For the spectra of SBPL fits without a peak, the break energy $E_\text{b}$ is used instead. For each spectrum, a triangle $\left\{(1, 1), (x_\text{left}, y_\text{left}), (x_\text{right}, y_\text{right})\right\}$ below the spectral curve is constructed in dimensionless space as
\begin{equation}
(x,y) =  \left\{
\begin{array}{ll}
(1, 1) \, , \\
(E_\text{left}/E_\text{p}, \nu F_\nu(E_\text{left})/\nu F_\nu(E_\text{p})) \, , \\
(E_\text{right}/E_\text{p}, \nu F_\nu(E_\text{right})/\nu F_\nu(E_\text{p}))
\end{array}
\right\} \, .
\end{equation}
The sharpness angle $\theta$ is computed in logarithmic space, under $(\log 1, \log 1)$ and between $\log x_\text{left}$ and $\log x_\text{right}$ (which we refer to as "triangle domain"). Thus, $\theta$ is an indication of the spectral sharpness and is independent of the actual position of $E_\text{p}$ (i.e., also independent of redshift). Figure~\ref{fig:triangle} illustrates how such a triangle can be constructed.

Similarly, we also construct the right-angled triangle $\left\{(1, 1), (x_\text{left}, y_\text{left}), (1, y_\text{left})\right\}$ below the spectral curve between $x_\text{left}$ and $x = 1$, and compute the left-hand side angle $\theta_\text{left}$ in logarithmic space, under $(\log 1, \log 1)$ and between $\log x_\text{left}$ and $\log 1$. Thus, in the limit of small $x_\text{left}$, $\theta_\text{left}$ becomes equivalent to a measure of the steepness of the low-energy power-law slope.

\subsection{Synchrotron emission models}
\label{subsect:sync}

The monochromatic flux of the synchrotron emission spectrum can be obtained, given the electron population $n_\text{e}$, as
\begin{equation}
\label{eqn:synchrotron}
F_\nu \propto \int_1^\infty n_\text{e}(\gamma_\text{e}) \, \mathcal{F}\left(\frac{\nu}{\nu_\text{e}}\right) \, \mathrm{d}\gamma_\text{e} \, ,
\end{equation}
where
\begin{equation}
\mathcal{F}(x) \equiv x \,  \int_x^\infty K_{5/3}(\epsilon) \, \mathrm{d}\epsilon \, ,
\end{equation}
in which $\gamma_\text{e}$ is the Lorentz factor of the electron, $\nu_\text{e}$ is the synchrotron frequency of the electron, and $K_{5/3}$ is the modified Bessel function of fractional order $5/3$. For a single electron, the synchrotron spectrum is simply proportional to $\mathcal{F}$.\footnote{This already assumes integration over emission direction \citep[see][Eqns. 6.29 - 31]{Rybicki86a}. If a single electron were viewed from a single angle, a sharper spectrum would mathematically result.} It can be shown that the limits of $\mathcal{F}(x)$ can be approximated by simpler analytical functions for $x \ll 1$ and $x \gg 1$ \citep[see, e.g.,][]{vanEerten09a}, for the ease of computation. Notice that Eqn.~(\ref{eqn:synchrotron}) either describes an instantaneously generated spectrum and 90 degrees pitch angle between magnetic field and electron velocity, or a situation where magnetic field and particle population remain unchanged.

Mathematically, the synchrotron emission spectrum of a single electron is the sharpest case. However, under realistic conditions, there is no reason to believe that the observed data originates from only one electron. Thus, it is more realistic to consider a Maxwellian population of electrons, since it is an efficient distribution of electron energies and sharper than typical non-thermal spectra. For a Maxwellian population of electrons with the temperature parameterized by the thermal Lorentz factor $\gamma_\text{th}$, we have
\begin{equation}
n_\text{e} \propto \left(\frac{\gamma_\text{e}}{\gamma_\text{th}}\right)^2 \exp\left(-\frac{\gamma_\text{e}}{\gamma_\text{th}}\right) \, ,
\end{equation}
and the Maxwellian synchrotron spectrum
\begin{equation}
\label{eqn:maxw}
F_\nu \propto \int_1^\infty \left(\frac{\gamma_\text{e}}{\gamma_\text{th}}\right)^2 \exp\left(-\frac{\gamma_\text{e}}{\gamma_\text{th}}\right) \, \mathcal{F}\left(\frac{\nu}{\nu_\text{e}}\right) \, \mathrm{d}\gamma_\text{e}  \, .
\end{equation}
Since $\nu_\text{e} \propto \gamma_\text{e}^2$, by changing the variable $\nu = \xi \nu_\text{th} \propto \xi \gamma_\text{th}^2$, it can be shown that
\begin{equation}
F_\nu \propto \gamma_\text{th} \, \xi^{\frac{3}{2}} \int_\frac{1}{\gamma_\text{th}}^\infty x^{-\frac{5}{2}} \exp\left(-\xi^\frac{1}{2} x^{-\frac{1}{2}}\right) \, \mathcal{F}(x) \, \mathrm{d}x \, ,
\end{equation}
which allows us to normalize the spectrum in units of $\xi$. Again, we note that Eqn.~(\ref{eqn:maxw}) represents one of the sharpest cases among synchrotron spectra for multiple electrons, but that the assumptions of a single temperature and magnetic field are still unrealistic. Observed emission will contain a mixture of these and lead to smoother spectra.

Another reasonable assumption for the electron population is a power-law distribution of the electron energies:
\begin{equation}
n_\text{e} \propto \gamma_\text{e}^{-p}: \gamma_\text{e} \geq \gamma_\text{m} \, ,
\end{equation}
and the synchrotron spectrum with population index $p$ is
\begin{equation}
\label{eqn:slow}
F_\nu \propto \int_{\gamma_\text{m}}^\infty \gamma_\text{e}^{-p} \, \mathcal{F}\left(\frac{\nu}{\nu_\text{e}}\right) \, \mathrm{d}\gamma_\text{e} \, ,
\end{equation}
where $\gamma_\text{m}$ is the minimum injection energy of the electron population. The $F_\nu$ spectrum can be solved as
\begin{equation}
F_\nu \propto \nu^{\frac{1-p}{2}} \int_0^{\frac{\nu}{\nu_\text{m}}} \left(\frac{\nu}{\nu_\text{e}}\right)^{\frac{p-3}{2}} \, \mathcal{F}\left(\frac{\nu}{\nu_\text{e}}\right) \, \mathrm{d}\left(\frac{\nu}{\nu_\text{e}}\right) \, ,
\end{equation}
where $\nu_\text{m}$ is the minimum injection frequency of the electron population. As with temperature, the observed spectrum will be smoother due to a mixture of $\nu_\text{m}$ values in the emission. If we substitute the approximation of $\mathcal{F}(x) \sim x^{1/3}$ for $x\ll1$, we can recover the 1/3 low-energy slope below $\nu_\text{m}$ for any value of $p$.

\begin{figure*}
\resizebox{\hsize}{!}
{\includegraphics[width = 18 cm]{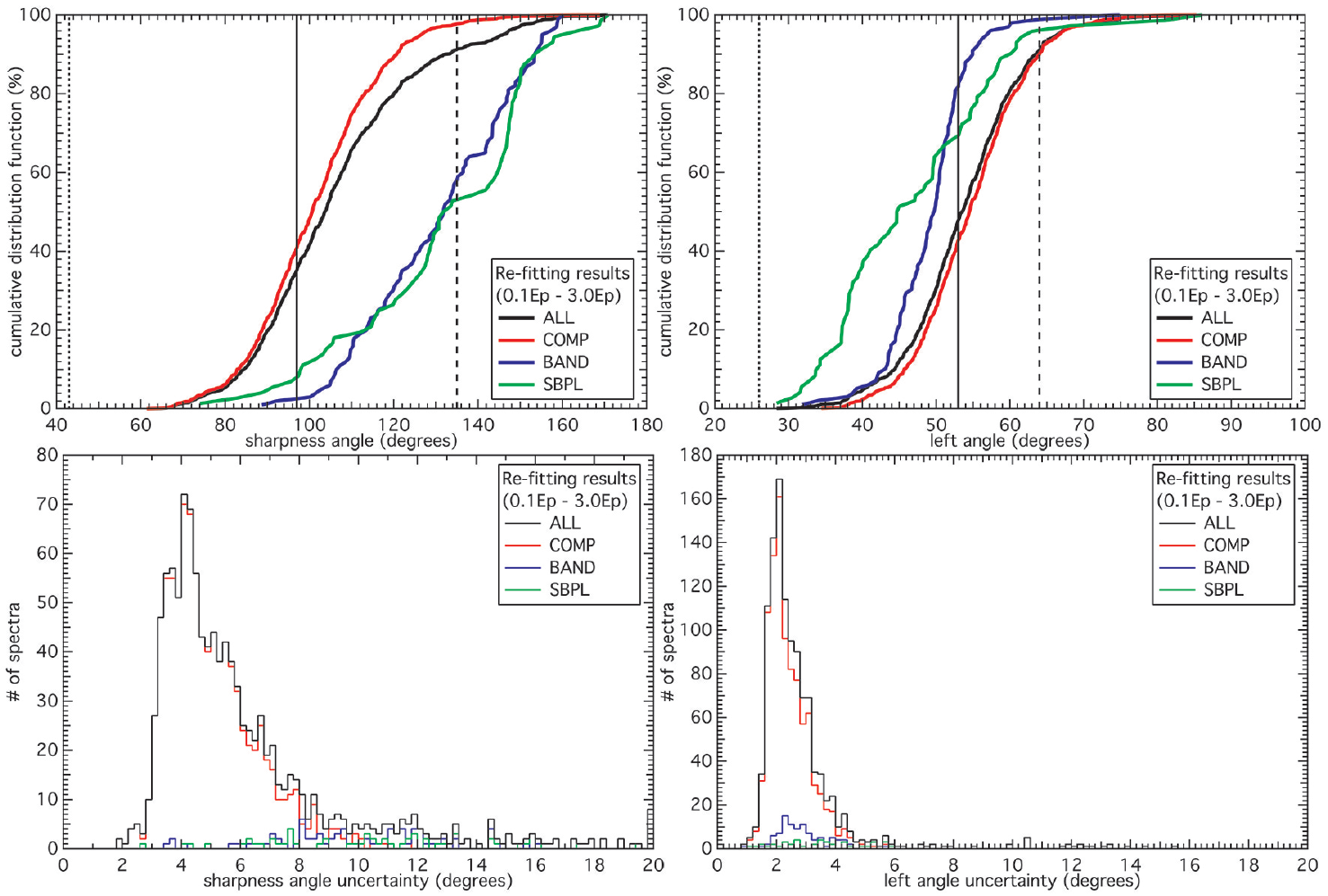}}
\caption{Left panels: Cumulative distribution functions of $\theta$ and distributions of $\sigma_\theta$. Right panels: Cumulative distribution functions of $\theta_\text{left}$ and distributions of $\sigma_\text{left}$. The limits of the normalized blackbody (dotted line), single-electron synchrotron (solid line), and synchrotron with a Maxwellian distribution function (dashed line) are overlaid. In the above legends, COMP represents the Comptonized model, BAND represents the Band function, SBPL represents the smoothly broken power law, and ALL represents the overall population (COMP + BAND + SBPL).}
\label{fig:main_dist}
\end{figure*}

In reality, electron cooling should exist, as more energetic electrons lose their energy faster due to radiative losses and cool down. One could consider, in addition to the minimum injection energy break $\nu_\text{m}$, the cooling break $\nu_\text{c}$ \citep[see, e.g., Fig.~1 in][]{Yu15a}. However, the ratio between $\nu_\text{m}$ and $\nu_\text{c}$ depends sensitively on assumptions on the shock micro-physics and fluid evolution. Additionally, the sharpness of the cooling break depends intrinsically on the distribution of electrons throughout the shock region, and has no local analog for a single electron population. For the purpose of this paper, it is sufficient to consider the case without cooling: quick evolution of electron energies due to cooling will smoothen the synchrotron spectrum. Therefore, any cooling synchrotron spectrum can never be sharper than Eqn.~(\ref{eqn:slow}).

For comparison, we also consider blackbody emission, which is given by
\begin{equation}
\label{eqn:bb}
F_\nu \propto \left[\frac{\nu^3}{\exp(h\nu/kT)-1}\right] \, .
\end{equation}

\section{Spectral sharpness results}
\label{sect:resu}

\begin{table}[!htbp]
\caption{Sharpness angle $\theta$ and left angle $\theta_\text{left}$ for various emission models. $^\text{a}$We note that if $p < 3$, $\nu F_\nu$ keeps on increasing monotonically.}
\label{tab:tab1}
\centering
\begin{tabular}{lcc}
\hline\hline
Emission models & $\theta$ (degrees) & $\theta_\text{left}$ (degrees) \\
\hline
Blackbody & 43 & 27 \\
Single-electron synchrotron & 97 & 53 \\
Maxwellian synchrotron & 135 & 64 \\
Synchrotron with $p = 2^\text{a}$ & 170 & 40 \\
Synchrotron with $p = 4$ & 128 & 56 \\
\hline
\end{tabular}
\end{table}

Figure~\ref{fig:main_dist} (left panels) shows the cumulative distribution functions (CDFs) of the sharpness angles $\theta$ and the distributions of the errors $\sigma_\theta$. The dotted, solid, and dashed black vertical lines indicate the values of $\theta$ for the normalized blackbody, single-electron synchrotron emission function,\footnote{For a single emission direction, $\theta$ and $\theta_\text{left}$ are 76 and 43 degrees respectively, for the polarization direction perpendicular to the projection of the magnetic field on the sky, and are 67 and 37 degrees respectively, in the parallel case. These values reflect the standard textbook results for single-electron emission prior to convolving with an electron distribution function \citep[see][]{Rybicki86a}.} and synchrotron emission function from a Maxwellian electron distribution, from left to right. These values are listed in Table~\ref{tab:tab1}. It is found that over 35\% of the spectra are inconsistent with single-electron synchrotron emission and 91\% are inconsistent with synchrotron emission from a Maxwellian electron distribution. The blackbody spectrum is found to be much sharper than any of the observed spectra.

The synchrotron emission function from a Maxwellian electron distribution produces one of the sharpest (i.e., narrowest) spectra (Sect.~\ref{subsect:sync}). The values of $\theta$ for the synchrotron emission function from a power-law electron distribution for $p=2$ and $p=4$ are also listed in Table~\ref{tab:tab1}. The spectrum for $p = 2$ was normalized by the peak position in the $F_\nu$ space, because for $p < 3$, $\nu F_\nu$ keeps on increasing monotonically. Notice also that the spectrum for $p = 4$ is of similar sharpness to the spectrum for Maxwellian (for $p \to \infty$, the spectrum would reduce to a single-electron synchrotron spectrum).

In principle, $\sigma_\theta$ should be propagated directly from the errors on the observed photon counts, since the counts are independent of the choice of the fitting models. However, the spectral peak can only be found and the flux can only be normalized when the counts are convolved with a model (COMP, BAND, etc.) and the response matrices, through RMFIT. Therefore, we compute $\sigma_\theta$ by performing Monte-Carlo simulations using the errors of the re-fit model parameters. First, we extract the $1\sigma$ errors from the RMFIT results. Because the errors on model fit parameters $\alpha$ and $\beta$ (see Appendix~\ref{app:fitfunc}) are not necessarily Gaussian, we then randomly draw new values of $\alpha$ and $\beta$ from a uniform probability function sharing the same $1\sigma$ error, and we re-compute $\theta$. This process is repeated 1,000 times for every spectrum. We then take the $1\sigma$ width of the resulting $\theta$ distribution and average over left and right $1\sigma$ values. We note that our method generates the most conservative values of $\sigma_\theta$, since the uniform probability function has the largest standard deviation.

As shown in the bottom left panel of Fig.~\ref{fig:main_dist}, the resulting distribution of $\sigma_\theta$ has a median around 5 degrees. This is too small to affect our conclusions. However, we note that $\sigma_\theta$ for BAND and SBPL can be systematically larger than those for COMP, because the high-energy tail of COMP is an exponential cutoff with no parameter dependence. Therefore, $\alpha$ has very little effect on the right-hand-side spectrum for COMP fits (see Eqn.~\ref{eqn:comp}), and $\sigma_\theta$ of COMP may be under-estimated.

Because the fit results for BAND and SBPL fits are distributed over a wide range and have larger angles and errors than COMP, it is of interest to look separately at the low-energy left-hand-side angles $\theta_\text{left}$. This way we can explore how both sides contribute to the total curvature and shape our results. Also, $\theta_\text{left}$ is unaffected by the transition from photon counts to upper limits that sometimes already occurs slightly below $3.0E_\text{p}$ on the right-hand-side. We therefore show, in the right panels of Fig.~\ref{fig:main_dist}, the CDFs of the low-energy left-hand-side angles $\theta_\text{left}$ (i.e., the angle under $(\log 1, \log 1)$ and between $\log x_\text{left}$ and $\log 1$) and the distributions of their errors $\sigma_\text{left}$.

The top right panel of Fig.~\ref{fig:main_dist} shows that if one compares $\theta_\text{left}$ instead of $\theta$, the overall fraction inconsistent with single-electron synchrotron increases to 48\%, and the overall fraction inconsistent with Maxwellian synchrotron is also 91\%. Therefore, even when only the left-hand-side of the spectral peak (or break) is considered, the same conclusions can be drawn, and the errors remain sufficiently small not to affect the final result. The distributions of the errors on $\theta_\text{left}$ are similar for the different fit functions, suggesting that their values are not merely driven by the curvature of the fit function itself. In addition, it shows that the low-energy curvature is the main cause of the violation of any synchrotron emission model, and that the upper limits in the high-energy side could harden the high-energy power laws of BAND and SBPL, which make the spectral shape less sharp.

Of our 1,113 spectra, 35\% violate the synchrotron line-of-death (i.e., $\alpha > -2/3$), higher than the 20\% observed by \citet{Gruber14a} in their peak-flux "\textit{P}" spectra sample. This implies that a large number of spectra are still consistent with the line-of-death. However, we find that in the 65\% of spectra that do not violate the line-of-death, 92\% of them violate the Maxwellian limit (i.e.,  $\theta < 135$~degrees) given in this paper. This shows that the sharpness angle method can identify many more spectra that are consistent with the line-of-death but are still sharper than what the synchrotron theory predicts. By contrast, of the 35\% of spectra that violate the line-of-death, only 10\% of them do not violate the Maxwellian limit.

\begin{figure}
\resizebox{\hsize}{!}
{\includegraphics{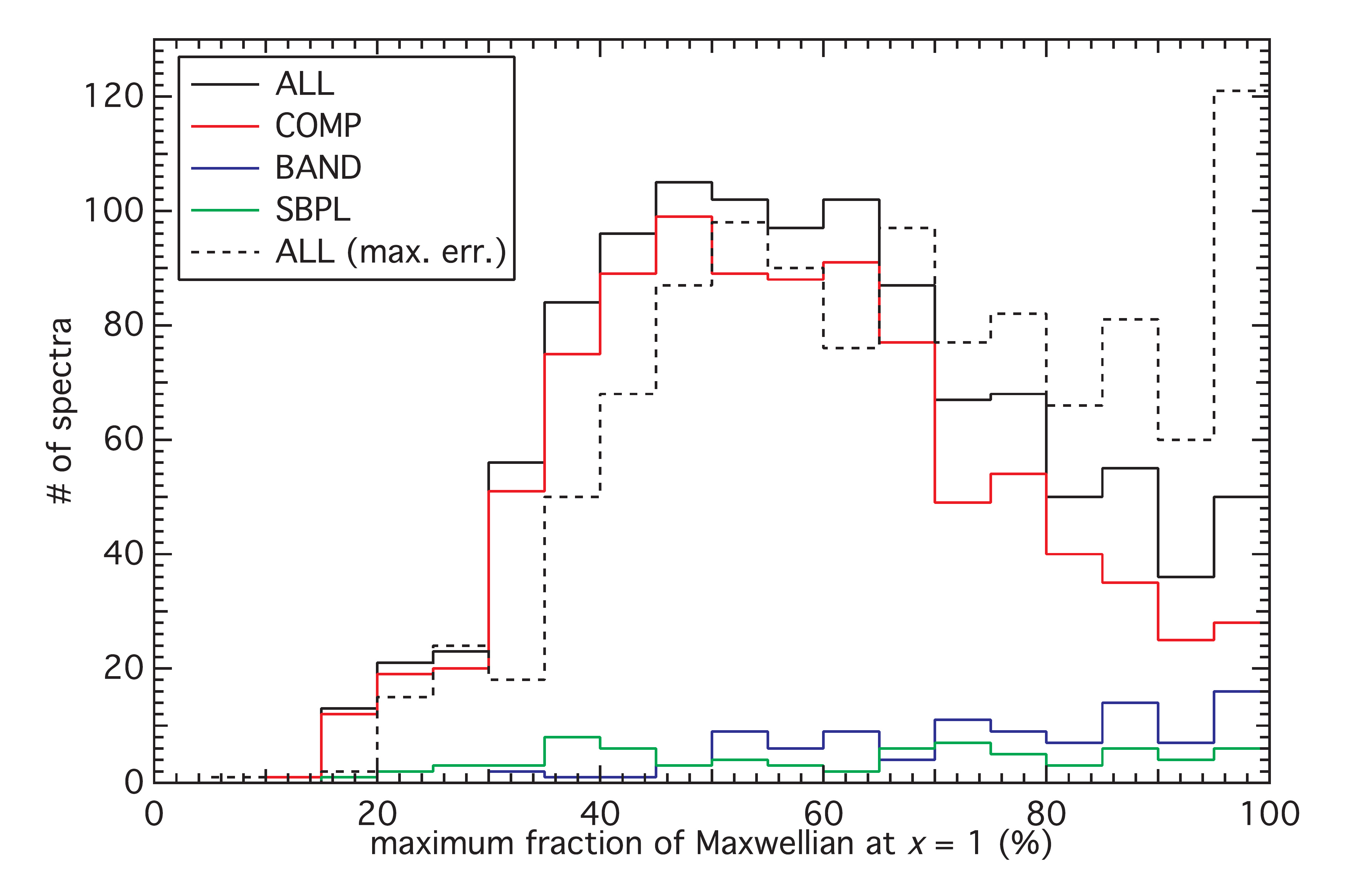}}
\caption{Distribution of the maximum fraction contributed from the Maxwellian synchrotron function at $x = 1$. The solid histograms represent the distributions using the best-fit model parameters, while the dashed histogram shows the minimum allowed sharpness by the uncertainties from the best-fit parameters. Spectra with 100\% at $x = 1$ are accumulated in the last bin.}
\label{fig:max_frac}
\end{figure}

\begin{figure}
\resizebox{\hsize}{!}
{\includegraphics{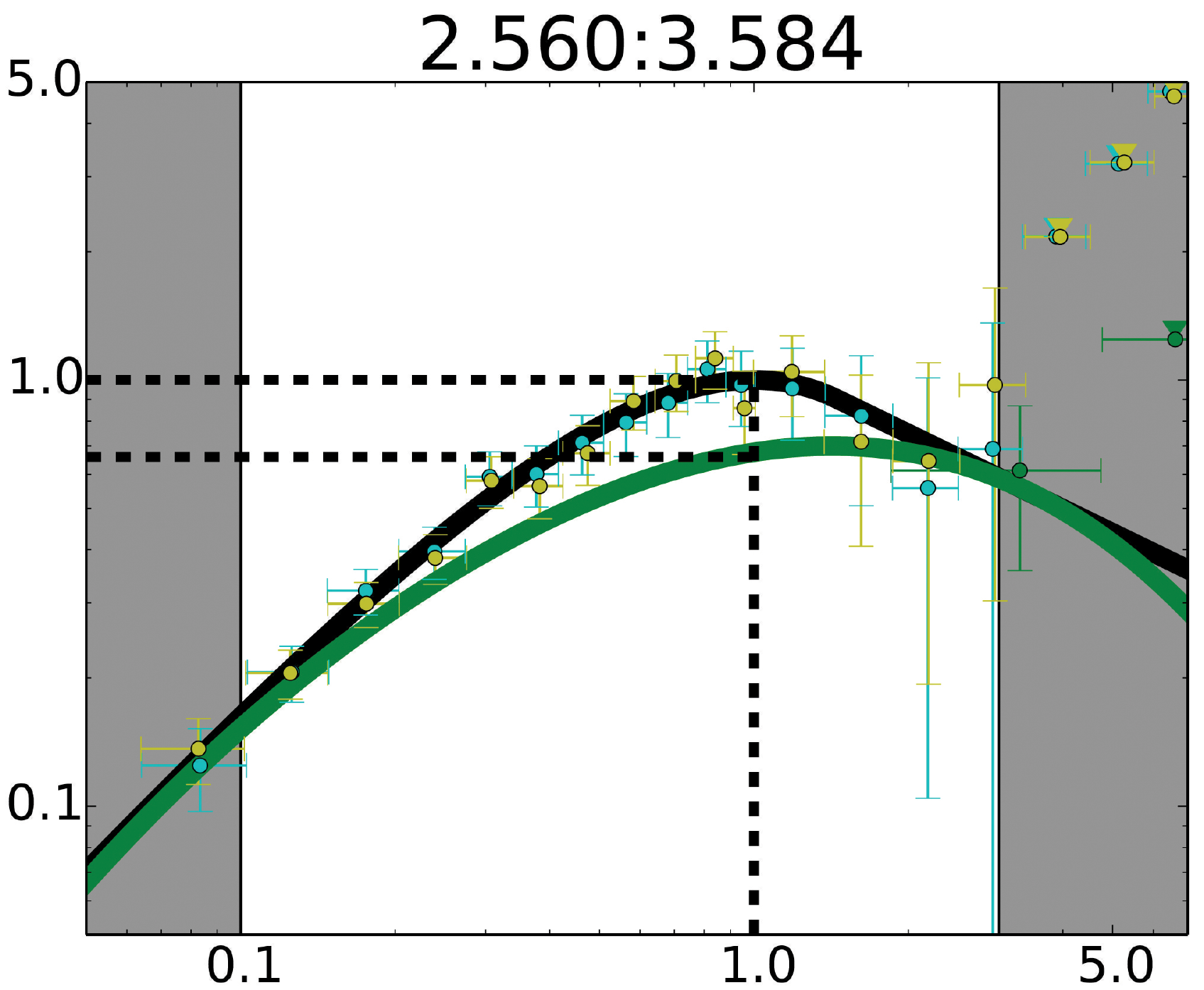}}
\caption{Example spectrum taken from GRB 101014.175 (2.560 - 3.584~s), showing the maximum contribution to the best-fit model by the Maxwellian synchrotron function, at $x = 1$. The normalized Maxwellian synchrotron (green curve) and the best-fit model (black curve) overlaid. The black dashed lines show the peak position of the best fit model and the relative normalized flux levels. In this particular spectrum, the Maxwellian fraction is about 65\% at $x = 1$. The shaded regions show the boundaries $x_\text{left} = 0.1$ and $x_\text{right} = 3.0$. Deep green data points are from the BGO detector and the others are from the NaI detectors. Triangles represent upper limits. For display purpose, the bin size has been increased by a factor of 5 - 10 relative to the standard bin size.}
\label{fig:minimise}
\end{figure}

\begin{figure*}
\resizebox{\hsize}{!}
{\includegraphics[width = 18 cm]{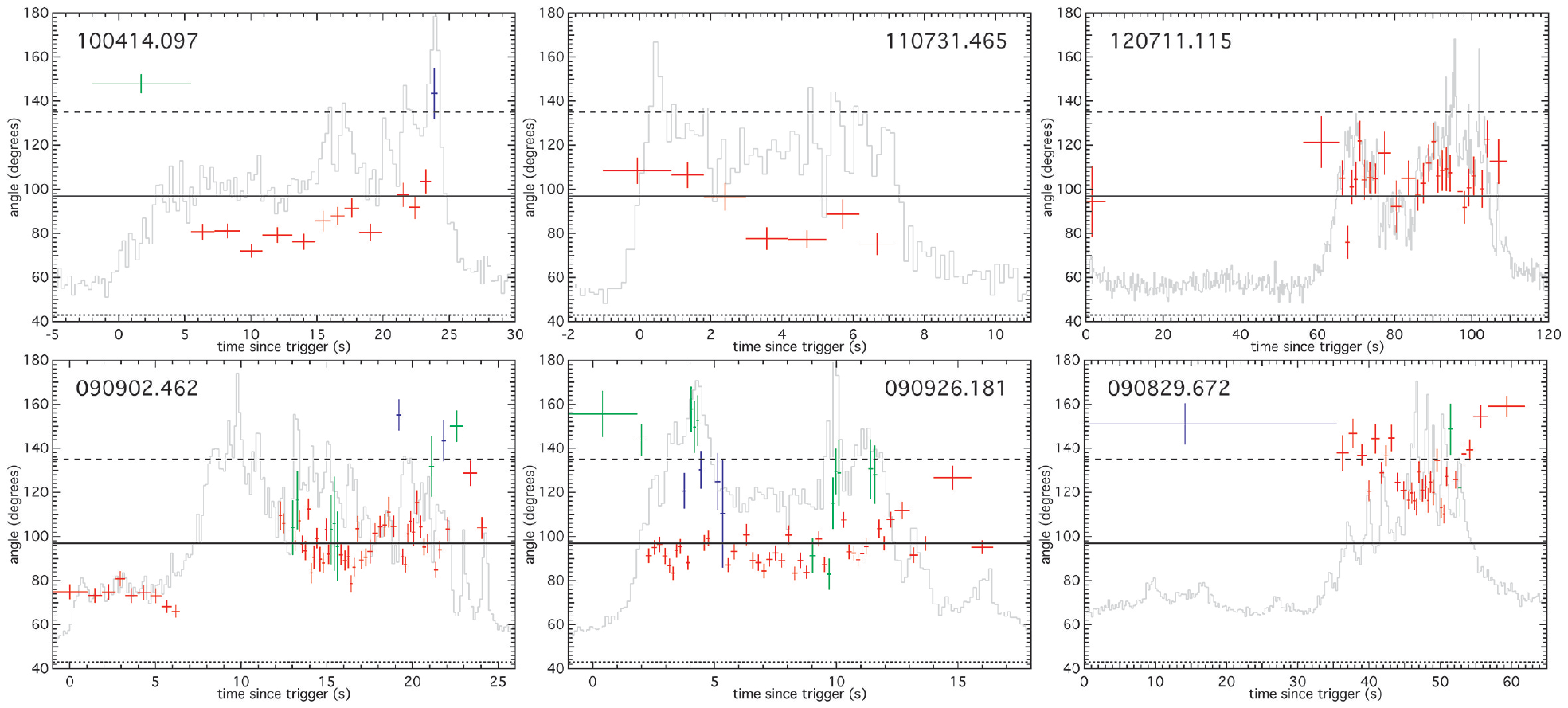}}
\caption{Six examples of evolutionary trends of $\theta$. Red, blue, or green color indicates that the best-fit model is COMP, BAND, or SBPL, respectively. The light curves are overlaid in arbitrary units. The limits of the normalized blackbody (dotted line), single-electron synchrotron (solid line), and synchrotron emission from a Maxwellian electron distribution (dashed line) are overlaid.}
\label{fig:lcs}
\end{figure*}

\begin{figure*}
\resizebox{\hsize}{!}
{\includegraphics[width = 18 cm]{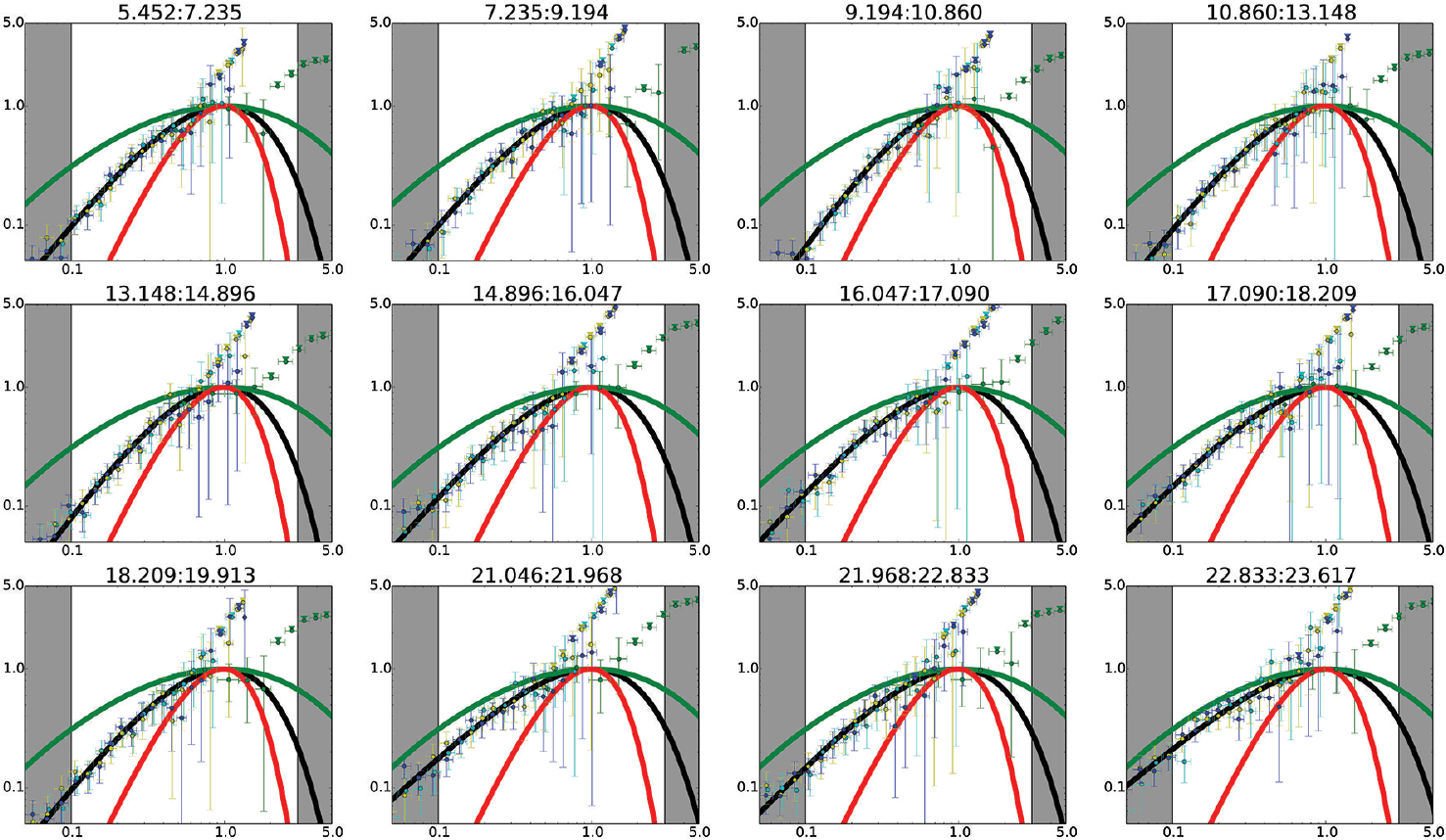}}
\caption{Spectral evolution of GRB 100414.097 with the normalized blackbody (red), Maxwellian synchrotron (green), and the best-fit model (black) overlaid. Time evolves from top left to bottom right, and the time since trigger is labeled at the top of each snapshot spectrum, in units of seconds. The peaks of the models are all normalized to $(x,y)=(1,1)$. Data points and the shaded regions are plotted as described in Fig.~\ref{fig:minimise}. For display purpose, the bin size has been increased by a factor of 5 relative to the standard bin size.}
\label{fig:spec_evo}
\end{figure*}

Since our results indicate that the synchrotron model alone cannot explain most of the prompt spectra, we can ask the question, if synchrotron emission is still one of the mechanisms that contributes to the observed peak flux, how much at most can it realistically contribute? In Fig.~\ref{fig:max_frac}, we show the distribution of the maximum peak flux contributed from the Maxwellian synchrotron function. For the spectra that do not have a peak, we compute this value at the spectral break. A sample spectrum from GRB 101014.175\footnote{In this paper, the names of the bursts are given according to the \textit{Fermi} GBM trigger designation that is assigned for each new trigger detected. The first 6 digits indicate the year, month, and day of the month, and the last 3 digits indicate the fraction of the day. For more details, please see the online \textit{Fermi} GBM burst catalog at \url{http://heasarc.gsfc.nasa.gov/W3Browse/fermi/fermigbrst.html}} is plotted in Fig.~\ref{fig:minimise}. The normalized Maxwellian synchrotron function was shifted vertically and horizontally until the distance between its value at $x=1$ and the peak of the fit model is minimized. The advantage of evaluating this value at the peak of the fit model is that it is energy domain independent. For the spectra of SBPL without a peak, the break energies $E_\text{b}$ are used instead. It is found that the Maxwellian can only contribute up to $58_{-18}^{+23}$\% of the peak flux (solid histogram). Even if the minimum sharpnesses (i.e., the broadest) allowed by the uncertainties in the best-fit parameters are considered, this only slightly increases to $68_{-23}^{+23}$\% (dashed histogram). Again, we caution that these synchrotron spectra represent a limiting case of high sharpness, relative to that expected from a distribution of temperatures and magnetic field strengths and a rapidly evolving particle population. In that sense, 58\% indicates an upper limit.

\subsection{Spectral evolution}

We now consider the sequence of spectra within bursts. We select and plot in Fig.~\ref{fig:lcs} the evolution of $\theta$ for 6 example bursts, with the Maxwellian synchrotron limit and the observed light curves overlaid. It can be seen that $\theta$ exhibits various evolutionary trends:

(1) In GRB 100414.097 (top left panel), the spectrum becomes less sharp as time evolves. We also plot the spectra of this burst in Fig.~\ref{fig:spec_evo}, with the normalized blackbody (red), Maxwellian synchrotron (green), and the best-fit model (black) overlaid. The violation of the Maxwellian synchrotron function is clearly shown in this example, and $\theta$ increases with time. In the typical fireball model, $\theta$ is expected to increase with time due to, e.g., increasing collision radius and curvature effects.

(2) In GRB 110731.465 (top central panel), the opposite happens and the spectrum becomes sharper as time evolves.

(3) In GRB 120711.115 (top right panel), $\theta$ fluctuates between the limits of single electron and Maxwellian, without clear correlation to the observed light curve. We note that the first time bin at around the trigger time has a small $\theta$.

(4) In GRB 090902.462 (bottom left panel), $\theta$ remains approximately constant in the plateau during the first 7~s, and then increases to higher but fluctuating values (11 - 25~s). We note that during 7 - 11~s, the catalog best-fit model is the power law plus blackbody, in accordance with the finding of \citet{Abdo09a}. We did not compute the sharpness angle for this period of time because the blackbody is sharper than all synchrotron cases.

(5) In GRB 090926.181 (bottom central panel), the low emission level first time bin gives the largest $\theta$, which is consistent with what Maxwellian synchrotron emission predicts (in contrast to GRB 090902.462), and $\theta$ then decreases and fluctuates around the value of the single-electron limit. It increases again in the penultimate time bin to a value marginally consistent with the Maxwellian limit, and then drops again to the single-electron limit.

(6) GRB 090829.672 (bottom right panel) has the largest fraction of spectra consistent with a Maxwellian synchrotron explanation (13 out of 32 spectra, 40\%). Similar to GRB 090926.181, it combines a large value of $\theta$ with a low emission level. During the main emission pulses between 35 - 55~s, $\theta$ decreases below the Maxwellian limit and then increases again to values above the Maxwellian limit.

\begin{figure*}
\resizebox{\hsize}{!}
{\includegraphics[width = 18 cm]{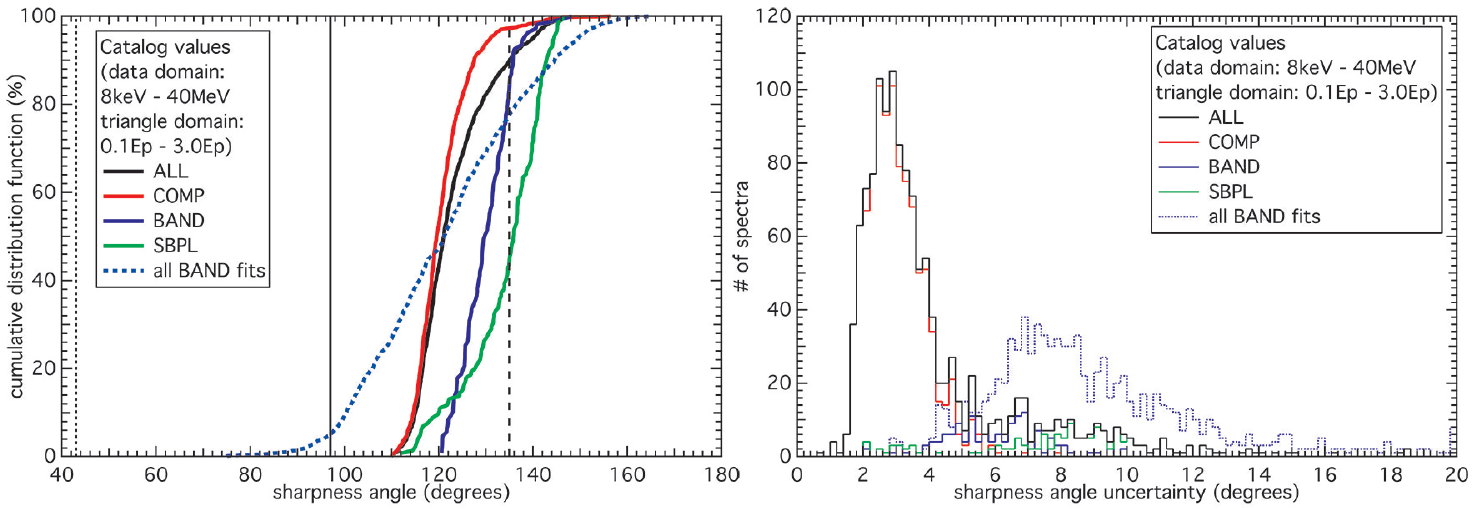}}
\caption{Left panel: Cumulative distribution functions of $\theta$ for the catalog full energy domain fits. Right panel: Distributions of $\sigma_\theta$. The limits of the normalized blackbody (dotted line), single-electron synchrotron (solid line), and synchrotron emission from a Maxwellian electron distribution (dashed line) are overlaid. In the above legends, COMP represents the Comptonized model, BAND represents the Band function, SBPL represents the smoothly broken power law, and ALL represents the overall population (COMP + BAND + SBPL). The blue dotted line and histogram show fit results if all spectra are fit using BAND, provided that they are converged fits, but not necessarily the best fit when compared to other models.}
\label{fig:consistency}
\end{figure*}

These bursts are chosen to show the variety of evolutionary trends in $\theta$: gradual increase, gradual decrease, fluctuation between the single-electron and Maxwellian limits, small $\theta$ during low emission level and large $\theta$ during high emission level, large $\theta$ during low emission level and small $\theta$ during high emission level, and decrease from above the Maxwellian limit followed by an increase again to above the Maxwellian limit.

\section{Consistency checks}
\label{sect:consistency}

\subsection{Choices of the fitting models}
\label{subsubsect:modelchoice}

It is observed that over 66\% of the time-resolved catalog best-fit models are COMP. This indicates that most of the observed spectra are indeed sharper than BAND or SBPL would predict. The same statistical behavior was also observed in the GBM time-integrated spectral catalogs \citep{Goldstein12a,Gruber14a} and the BATSE time-integrated spectral catalogs \citep{Kaneko06a,Goldstein13a}.

We show in Fig.~\ref{fig:consistency} the CDFs of $\theta$ and the distributions of $\sigma_\theta$ from the catalog best fits, evaluated using the full data domain of (8~keV, 40~MeV) and triangle domain of $(x_\text{left}, x_\text{right}) = (0.1, 3.0)$. It is observed that the catalog best fits produce results similar to the re-fits.

We note that COMP is inherently an exponential cutoff model, while BAND and SBPL are power laws joined by a peak or break energy. This intrinsic difference between the fit functions motivates us to explore the fit results if all spectra are fit using the Band function. Therefore, we further plot in Fig.~\ref{fig:consistency} the distributions of $\theta$ and $\sigma_\theta$ using the catalog BAND fits ("all BAND"), provided that it is a converged fit with a peak (or break) energy in the $\nu F_\nu$ space, but not necessarily the best fit when compared to other models. We find that it gives larger $\theta$ and $\sigma_\theta$. This indicates that when the Band function is applied to all spectra, the values of $\theta$ can be over-estimated due to larger uncertainties. Nevertheless, even in the all-BAND approach, 77\% of spectra are sharper than the Maxwellian synchrotron limit.

\begin{figure}
\resizebox{\hsize}{!}
{\includegraphics{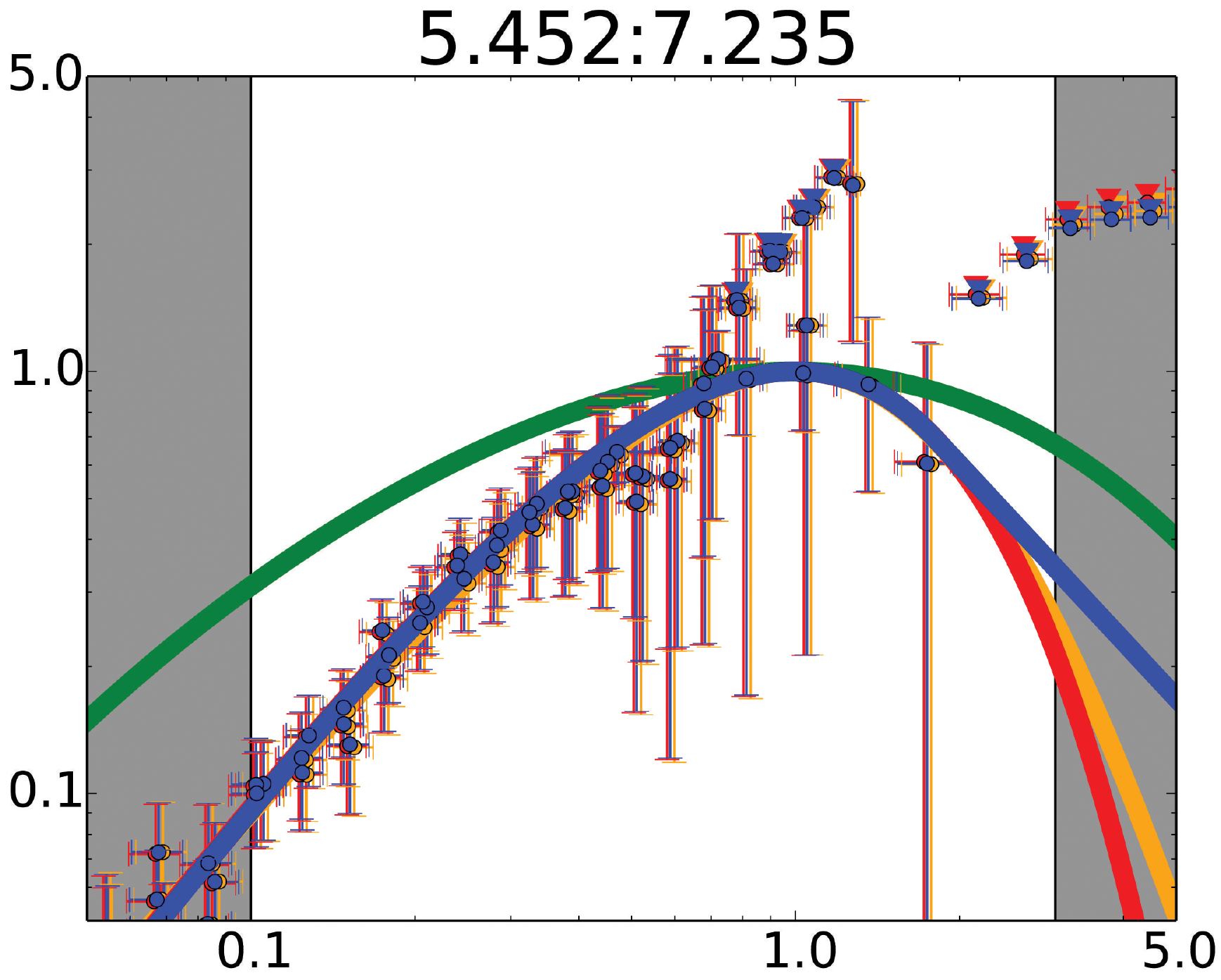}}
\caption{Comparison of the convolved data points and the respective convolving model curves for a sample spectrum taken from GRB 100414.097. The red curve and data points are obtained from the COMP fit, the blue ones are from the BAND fit, and the orange ones are from the SBPL fit with the break scale $\Delta$ allowed to vary. The Maxwellian synchrotron function is also overlaid (green). For display purpose, the bin size has been increased by a factor of 5 relative to the standard bin size.}
\label{fig:different_models_data}
\end{figure}

\begin{figure}
\resizebox{\hsize}{!}
{\includegraphics{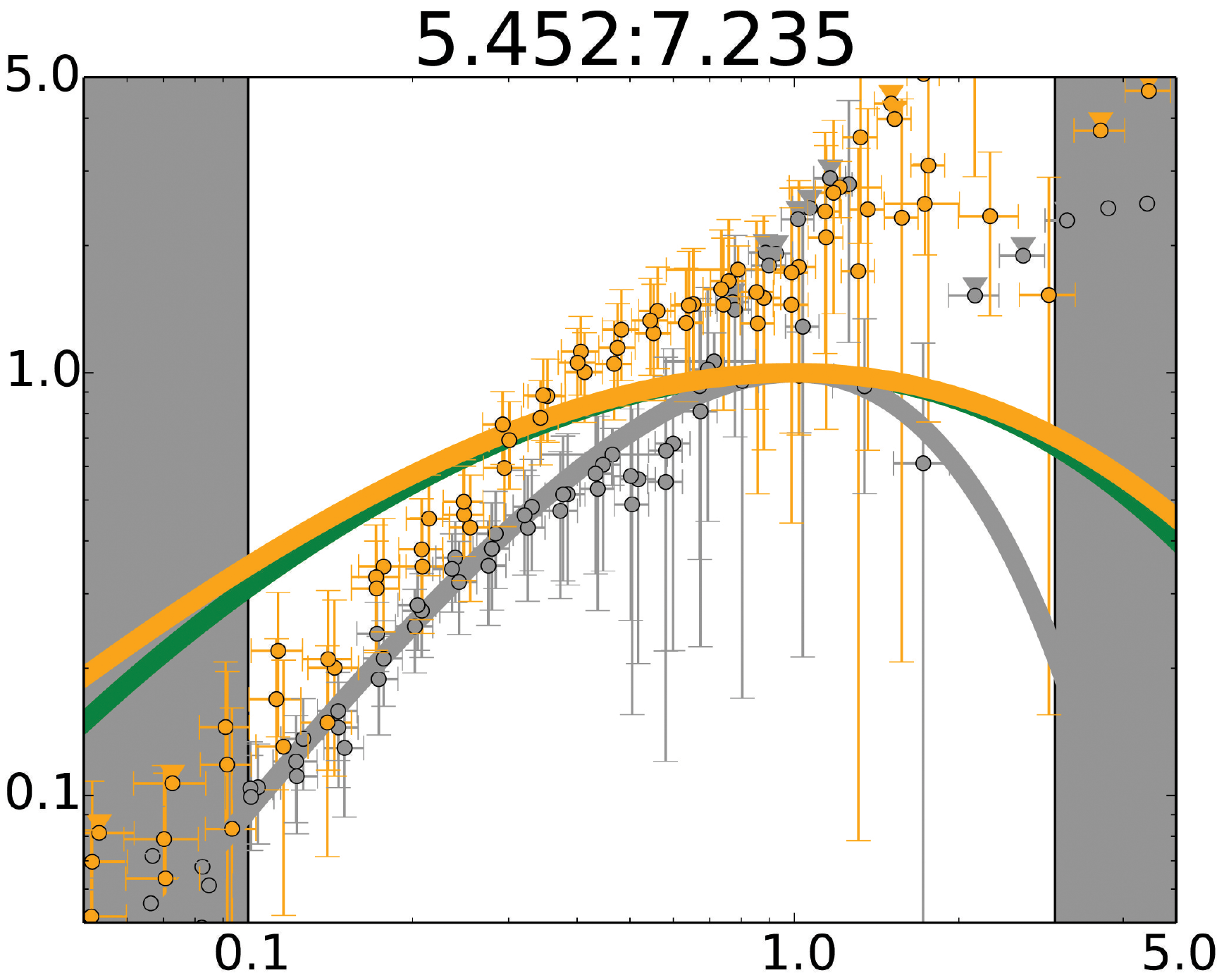}}
\caption{Comparison of the convolved data points and the respective convolving model curves for the same spectrum of Fig.~\ref{fig:different_models_data}. The orange curve and data points are obtained from the Maxwellian-SBPL fit, and the grey ones are from the COMP fit. The Maxwellian synchrotron function is also overlaid (green). For display purpose, the bin size has been increased by a factor of 5 relative to the standard bin size.}
\label{fig:Max-SBPL}
\end{figure}

The models of COMP, BAND, and SBPL have been extensively tested over the years and are found to provide good fits to data \citep[e.g.,][]{Kaneko06a}. In Fig.~\ref{fig:different_models_data}, we show the comparison of the convolved data points and the respective convolving model curves for an illustrative sample spectrum taken from GRB 100414.097 (see also Fig.~\ref{fig:spec_evo}). The red curve and data points are obtained from the COMP fit (CSTAT/dof\footnote{The modified Cash Statistics \citep{Cash79a}, Caster C-Statistics, per degrees of freedom.} = 301.66/285), the blue ones are from the BAND fit (CSTAT/dof = 301.61/284), and the orange ones are from the SBPL fit (CSTAT/dof = 301.68/283) with the break scale $\Delta$ allowed to vary (see Eqn.~\ref{eqn:sbplpara}). The fit functions start to diverge when extrapolated outside the data domain, but the data points of different convolving models coincide almost exactly, even when $\Delta$ is left as a free parameter. This indicates that these empirical functions provide good descriptions of the observed data, justifying our choices of models to obtain the spectral sharpness angles. However, we note that the fit parameters of SBPL become unconstrained for a varying $\Delta$, which indicates degeneracy in the parameter space. Therefore, we follow the catalogs and fix $\Delta = 0.3$.

Figure~\ref{fig:Max-SBPL} repeats (in grey) the data points convolved with the best fit function (COMP) from Fig.~\ref{fig:different_models_data} and shows a comparison with a SBPL that mimics a Maxwellian synchrotron function (shown in orange, overlaid on the original Maxwellian in green). First we separately fit a SBPL to the Maxwellian synchrotron model in order to obtain a curve that can be used directly in RMFIT. Then we fit this Maxwellian-SBPL function to the data by fixing the fit parameters except for the normalization factor $A$. This demonstrates how the data points can shift under convolving with a strongly differing fit function. Even though the data points shift, the resulting fit is significantly worse (CSTAT/dof = 558.00/287) than the COMP fit. This is consistent with work by \citet{Burgess14a}, who directly convolved synchrotron emission spectra with photon counts and found CSTAT values differing typically by hundreds relative to best fit curves.

\subsection{Choice of the data domain}
\label{sect:datadomain}

\begin{figure}
\resizebox{\hsize}{!}
{\includegraphics{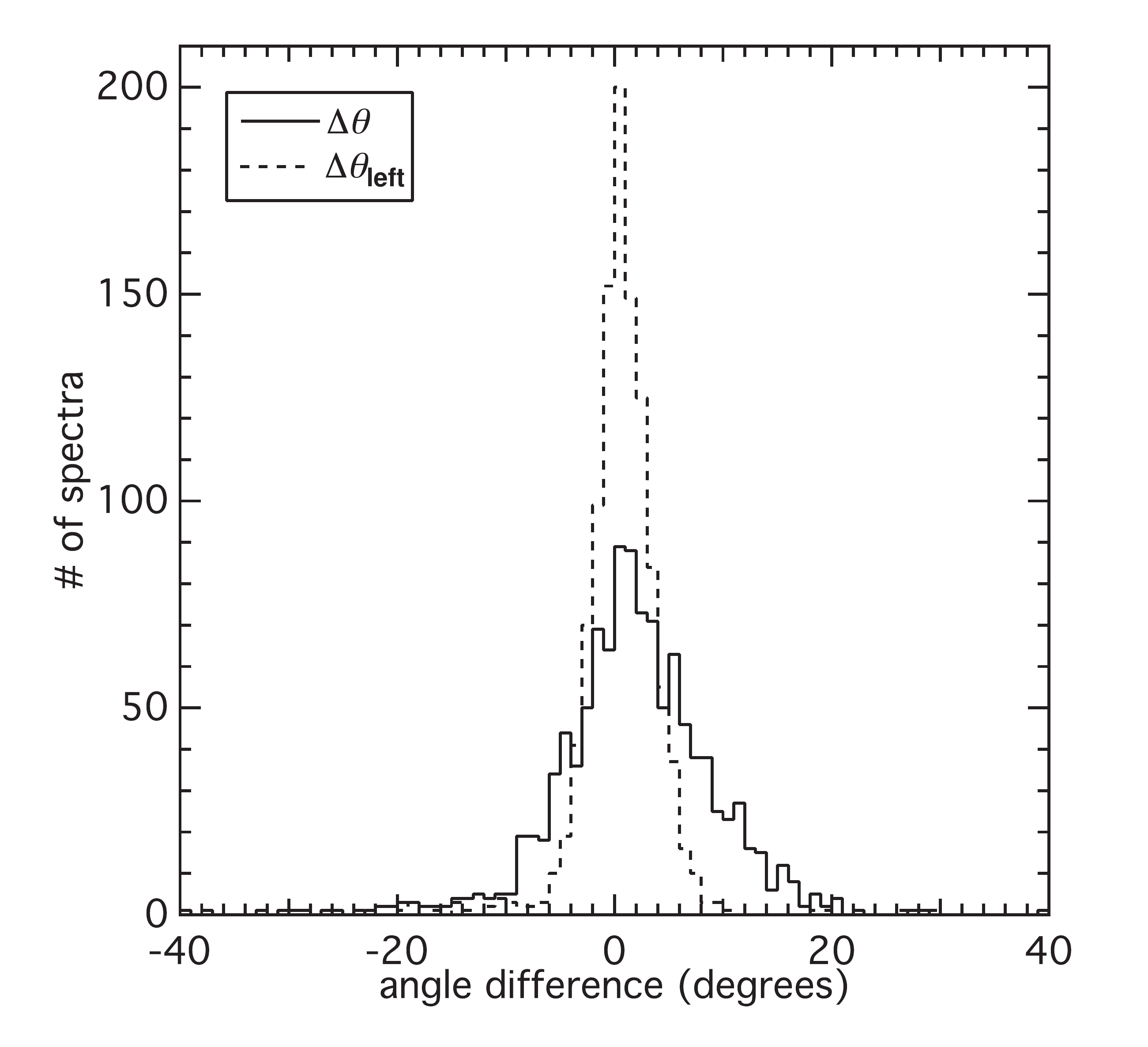}}
\caption{Distributions for $\Delta \theta = \theta^\text{catalog} - \theta^\text{re-fit}$ (solid histogram) and $\Delta \theta_\text{left} = \theta_\text{left}^\text{catalog} - \theta_\text{left}^\text{re-fit}$ (dashed histogram). The values of $\Delta \theta$ and $\Delta \theta_\text{left}$ are normally distributed with medians at 1.7 and 0.6 degrees, respectively.}
\label{fig:difference}
\end{figure}

A key distinction between the current work and others, is that in this work, we want to obtain a mathematical description of the \textit{peak or break curvature} rather than of the whole spectrum. For this reason, we need to test whether our results hold up under a change in the data domain. The considerations when choosing the data domain size are (1) we want to have as many data points as possible, while (2) we do not want to include data too far away from the spectral peak, which could introduce extra curvature effects on the low-energy end and too many upper limits on the high-energy end that might pull the best fit function away from the data points near the peak or break, or shift the inferred peak or break itself. For smaller data domains, some of the fit parameters can be more weakly constrained than when the full energy domain is used. However, as we show in Sect.~\ref{sect:angleodmain}, the violation of the synchrotron emission model cannot be explained by the errors on the re-fit parameters.

In order to find the optimal data domain size, the re-fitting process is repeated using different values of $E_\text{right}$, for spectra of the few brightest bursts. We find that the first-upper-limit-data point of the BGO detector is typically at 1.5 - 3.5 times the value of $E_\text{p}$. Therefore, in order to minimise the effect due to the high-energy upper limits, $E_\text{right} = 3.0 E_\text{p}$ is adopted. The above checking process is again repeated with different values of $E_\text{left}$. It is found that data domains smaller than $(0.1 E_\text{p}, 3.0 E_\text{p})$ usually produce large uncertainties in the fits because the data are insufficient to define a definite functional shape.

In Fig.~\ref{fig:difference}, we show the differences in $\theta$ and $\theta_\text{left}$ between the catalog domain size (i.e., the full GBM energy domain from 8~keV to 40~MeV) and the data domain size mainly used in this paper (i.e., $0.1 E_\text{p}$ to $3.0 E_\text{p}$), for each spectrum. We find that $\Delta \theta$ and $\Delta \theta_\text{left}$ are normally distributed with medians at 1.7 and 0.6 degrees, respectively. This shows that while the extra curvature effects contributed by the data points on the flanks lead to a change in smoothness, the effect is small, and limiting the data domain size is not strictly necessary (this is also confirmed by Fig.~\ref{fig:consistency}).

\subsection{Choice of the triangle domain}
\label{sect:angleodmain}

\begin{figure*}
\resizebox{\hsize}{!}
{\includegraphics{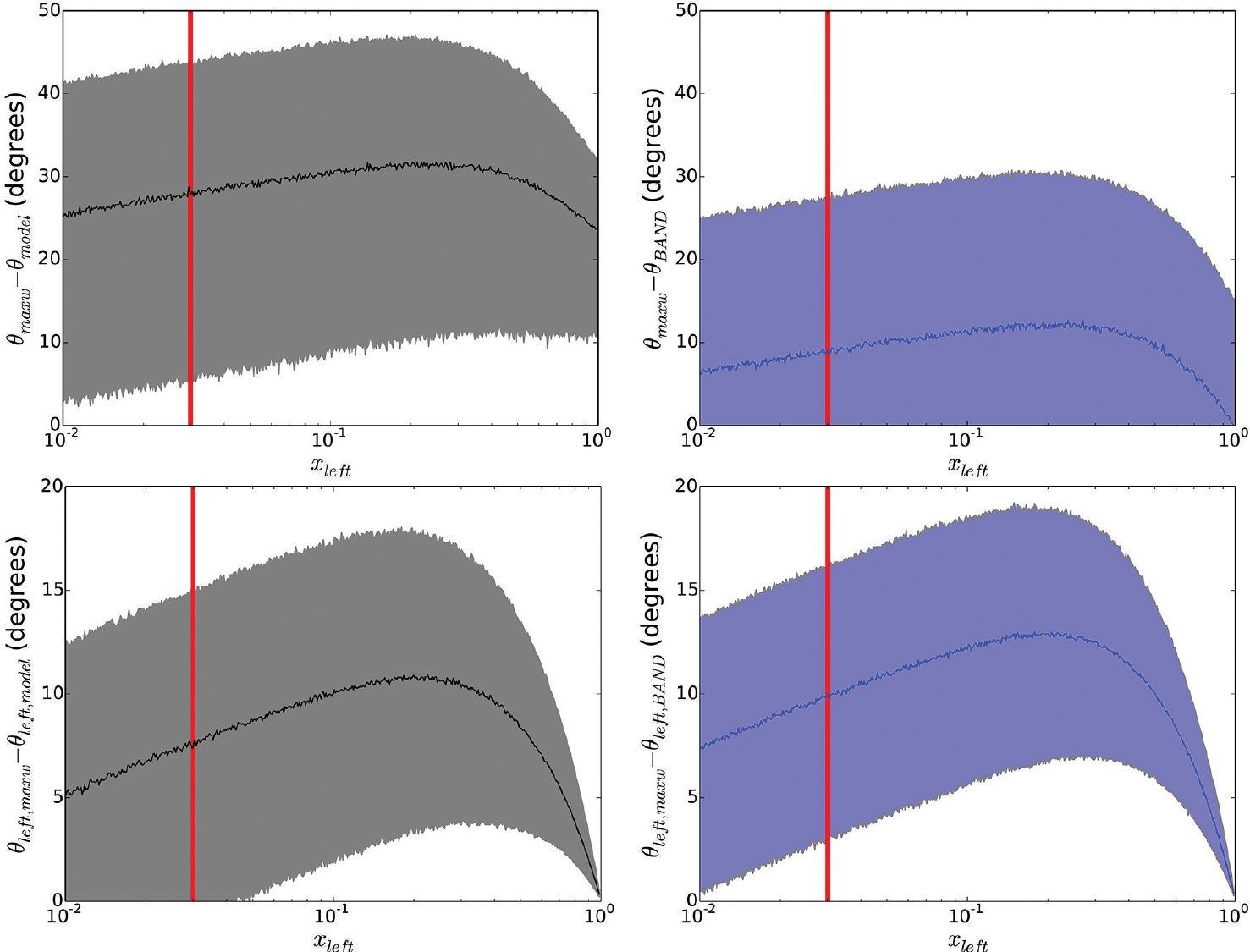}}
\caption{Top left panel: Difference between the sharpness angles of the Maxwellian synchrotron function and the re-fitting models, $\theta_\text{maxw}-\theta_\text{model}$, as a function of $x_\text{left}$. Top right panel: Same for $\theta_\text{maxw}-\theta_\text{BAND}$. Bottom left panel: Same for $\theta_\text{left,maxw}-\theta_\text{left,model}$. Bottom right panel: Same for $\theta_\text{left,maxw}-\theta_\text{left,BAND}$. The red vertical lines show the NaI detector limit of 8~keV$\sim0.03E_\text{p}$ for $E_\text{p} \sim 250$~keV. The shaded regions show the $1\sigma$ regions. See main text for details about the plots.}
\label{fig:angle_domain_checks}
\end{figure*}

Besides the re-fitting data domain, we also check the validity of the triangle domain used in the computation of $\theta$. There are 3 choices: (1) triangle domain $>$ data domain, (2) triangle domain $=$ data domain, and (3) triangle domain $<$ data domain.

Triangle domain $>$ data domain is obviously not statistically sound, because we have no knowledge of how the data behave outside the data domain. On the other hand, we need to find a balance between staying as close to the peak (or break) as possible and measuring a meaningful amount of curvature. As discussed in Sect.~\ref{sect:datadomain}, the choice of $x_\text{right}$ is already limited by the upper limits, so we concentrate on checking the consistency of $x_\text{left}$.

In Fig.~\ref{fig:angle_domain_checks}, we show the difference between the sharpness angles of the Maxwellian synchrotron function and the fitting models as a function of $x_\text{left}$ (top left panel) and that between the Maxwellian synchrotron function and the all-BAND fit results (i.e., those described in Sect.~\ref{subsubsect:modelchoice} and shown in Fig.~\ref{fig:consistency}). The same plots for the differences of $\theta_\text{left}$ are shown in the bottom panels. These plots are produced according to the procedure described below. First, a re-fit spectrum is randomly chosen. Second, we randomly draw new values of $\alpha$ and $\beta$ from a uniform probability function characterized by the $1\sigma$ errors of the spectrum, and a new value of $\theta$ is computed. Third, we repeat the first two steps 10,000 times and obtain the distributions for different values of $x_\text{left}$. We note that the plots have extrapolated below the NaI detector limit of 8~keV$\sim0.03E_\text{p}$ for $E_\text{p} \sim 250$~keV (close to the median $E_\text{p}$ time-resolved catalog value from Yu et al. in prep.), indicated by red vertical lines in Fig.~\ref{fig:angle_domain_checks}. The shaded bands show the $1\sigma$ region.

For triangle domain choices where the lower boundary of the shaded band lies above 0, the difference between data and synchrotron theory is the clearest. The plot therefore shows how $x_\text{left} = 0.1$ robustly leads to an unambiguous result. This is true for other choices of $x_\text{left}$ as well, as long as $x_\text{left} \gtrsim 0.05$. Fig.~\ref{fig:angle_domain_checks} also shows that, while setting $x_\text{left} = 0.3$ rather than 0.1 leads to a larger safety margin, the difference in actual angle is negligible. By extrapolating the triangle domain boundary $x_\text{left}$ to very small values, the long side of the triangle will eventually align with the left power-law asymptote. For any basic synchrotron spectrum, the left angle will then approach $\theta_\text{left} = \sin^{-1}(3/4) \approx 48.6$~degrees, corresponding to the well-known synchrotron line-of-death slope. Subsequently comparing this angle to one inferred from a best fit, therefore then becomes equivalent to testing for violation of the synchrotron line-of-death. We note however, that this analysis indicates a large error margin and extrapolating beyond the data domain.

\section{Theoretical implications}
\label{sect:disc}

Our results show that for most GRB prompt emission spectra, an explanation in terms of synchrotron radiation can be problematic. In the internal shocks of GRBs, a single-electron emission function is obviously non-realistic (as there must be multiple electrons in the outflow) and a Maxwellian population drawn from a single temperature is the limiting case. Even this limiting case is already too wide to fit most GRB time-resolved spectra.

\begin{figure}
\resizebox{\hsize}{!}
{\includegraphics{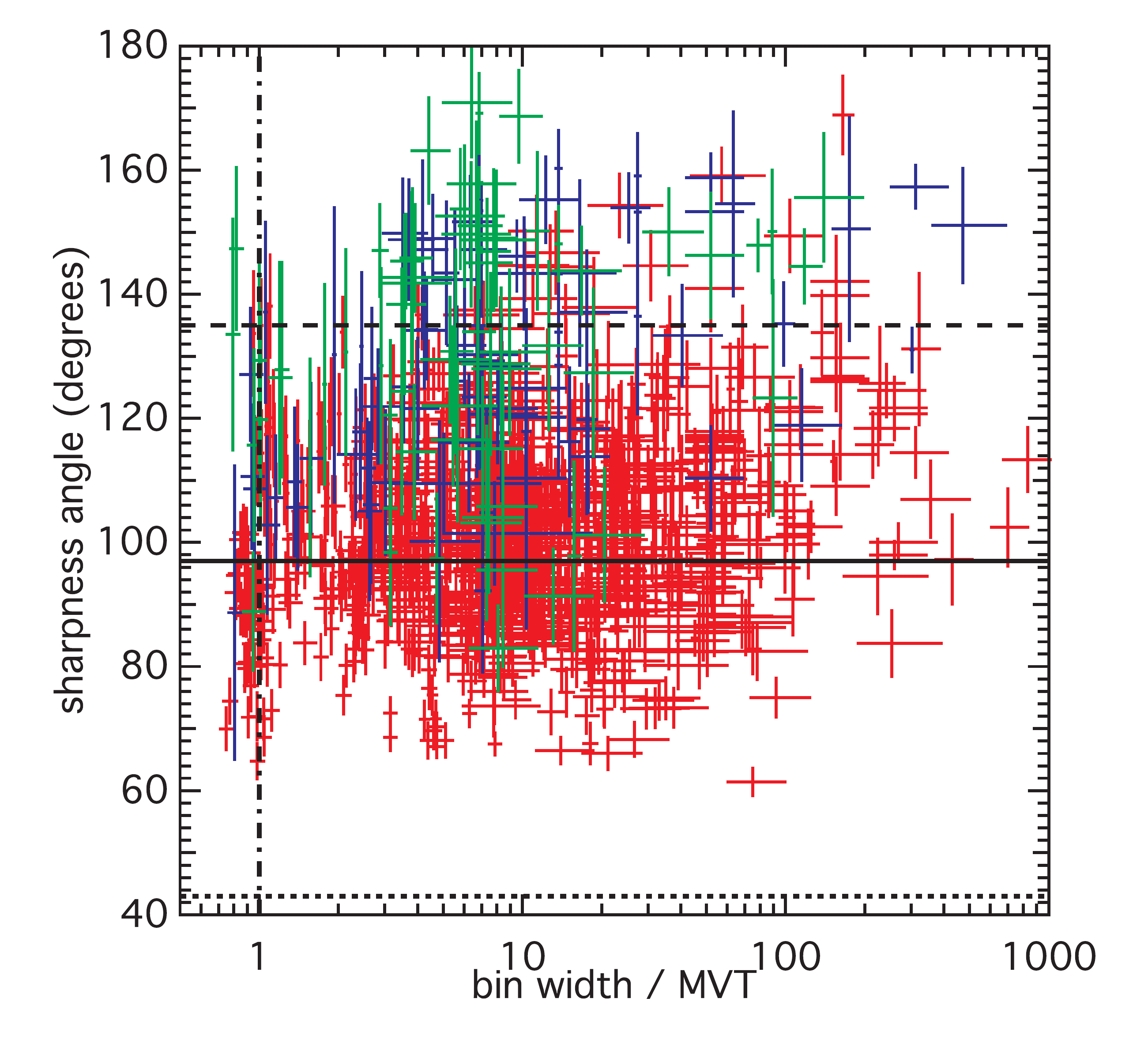}}
\caption{Sharpness angles plotted against the temporal bin widths per MVT. Red dat points show spectra best fit by COMP, blue by BAND, and green by SBPL. The vertical dash-dotted line shows where the bin width equals the MVT, only 4.4\% of data points are located to the left of the line. The horizontal lines show the limits of the normalized blackbody (dotted), single-electron synchrotron (solid), and synchrotron emission from a Maxwellian electron distribution (dashed).}
\label{fig:mvt}
\end{figure}

The minimum variability timescale \citep[MVT, e.g.,][]{Bhat13a,Golkhou14a} of the light curves is thought to be related to the actual dynamical timescale of the emission process. Therefore, if the temporal bin widths of our spectra are larger than the MVTs, then in the time-resolved catalog we are still looking at averaged time-resolved spectra that are less averaged than those in the time-integrated catalogs. In Fig.~\ref{fig:mvt}, we plot $\theta$ against temporal bin widths per MVT \citep[for the computational method of the MVT, see][]{Bhat13a}. It is observed that, in 1,064 spectra (49 spectra were excluded because they belong to bursts with no MVT due to bad or not enough GBM data), only 4.4\% of the spectra have bin width less than the MVT for the respective burst. This means that the problem for the synchrotron theory may be even more severe, since our spectra are smoothened already. However, this picture is complicated by the possibility that the MVT is time and energy dependent. \citet{Golkhou14a} use another method to compute the MVT, which is consistent with our method \citep{Golkhou15a}. The uncertainties of the MVT are worthy of independent studies which are beyond the scope of this paper.

For many years, the Band function has been assumed to be the appropriate mathematical description in most of the GRB prompt spectral studies. As already shown in Sect.~\ref{subsubsect:modelchoice}, the fact that most spectra are best fit by the Comptonized model (both time-integrated and time-resolved) shows that the high-energy tail of the prompt spectrum is actually sharper than a Band function would predict (i.e., maybe somewhere in between BAND and COMP). In a recent study using a subsample of GBM bursts which occurred in the \textit{Fermi} Large Area Telescope\footnote{The \textit{Fermi} Large Area Telescope is a pair production telescope covering the energy range from 20~MeV to 300~GeV.} \citep[LAT,][]{Atwood09a} field of view but remained undetected, \citet{Ackermann12a} showed that the Band function's $\beta$ (as obtained from GBM spectral fits) is too hard to be consistent with the LAT upper limits. All these results are indicating that the Band function can lead to incorrect interpretation of the data. To resolve this problem, there are at least two ways: (1) to invent another empirical mathematical function and then again try to interpret the parameters of this new function by physics; or (2) to fit the observed spectrum directly by physical models.

It is difficult to construct another empirical function which can improve upon the Band function, because it is already very simple in a statistical sense: it has only four parameters, and COMP has three. \citet{Yu15a} have shown that a triple power law with sharp breaks, in which the power-law indices have already been constrained according to the fast- or slow-cooling synchrotron models, could only perform as good as the Band function. They have found that in many cases even an extra blackbody is needed to describe the spectral curvature. Recently, more and more studies are being performed using physical fitting models \citep[e.g.,][]{Burgess11a,Burgess14a} and simulations under more realistic physical conditions, e.g., varying magnetic fields \citep[e.g.,][]{Uhm14a}. However, without knowledge of the emission process, it is difficult to formulate a sufficiently well-constrained fit function. Furthermore, there may be multiple emission mechanisms at work, the sum of which forms the observed prompt spectra.

The fitting results obtained using semi-empirical synchrotron models \citep[e.g.,][]{Yu15a} and physical synchrotron models \citep[e.g.,][]{Burgess11a,Burgess14a} show that extra thermal components are needed to fit the data. The resulting poor CSTAT values and systematic residual trends indicate that a pure non-thermal synchrotron emission function is inconsistent with the data at the peak or break energies, and thus cannot be the dominant process which contributes to the observed flux around this energy range. The distribution of spectral peak sharpness values that we report in this paper implies that any model based on standard synchrotron theory without additional radiative mechanisms will systematically struggle to capture the spectral curvature of the prompt emission. This will manifest itself in relatively poor CSTAT values and systematic trends in the fit residuals.

Recently, \citet{Axelsson15a} have shown that using the full-width-half-maximum measurement of GRB prompt emission spectra taken from the BATSE 5B GRB spectral catalog \citep{Goldstein13a} and 4-years \textit{Fermi} GBM GRB time-integrated spectral catalog \citep{Gruber14a}, a significant fraction of bursts (78\% for long and 85\% for short GRBs) could not be explained by a Maxwellian population-based slow-cooling synchrotron function. Our results show that using the time-resolved spectra this violation is actually more severe, with over 91\% of spectra obtained from long bursts violating the Maxwellian synchrotron function drawn from a single temperature, which is already a limiting case.

As can be seen from Fig.~\ref{fig:spec_evo}, it is obvious that a small number of Planck functions are not enough to reconstruct the observed spectral shape. From the observational point-of-view, fitting many blackbodies (with many parameters) is statistically meaningless, although maybe a sufficiently simple function describing a continuum of temperatures can be formulated. On the theoretical side, simple photospheric models also show difficulties in explaining the observed data. For example, early theoretical studies of a pure thermal origin of GRB prompt emission, such as from freely expanding photospheric outflows with no baryonic matter or magnetic field \citep{Goodman86a,Paczynski86a}, have shown difficulties in explaining the shape of the prompt emission phase and the two evolutionary trends of $E_\text{p}$ \citep[i.e.. hard-to-soft evolution and intensity tracking, see, e.g.,][]{Ford95a}. Recent studies \citep[e.g.,][]{Peer06a,Giannios08a,Peer11a,Ryde11a,Vurm11a,Lazzati13a} suggested that the Band function can be reconstructed from a thermal model. However, \citet{Deng14a} claim that the hard-to-soft evolution of $E_\text{p}$ is difficult to reproduce under natural photospheric conditions.

A frequently discussed alternative to the baryonic composition of the jets in GRBs is a magnetically, or Poynting flux, dominated jet \citep{Thompson94a,Drenkhahn02a,Lyutikov03a}. In this scenario, the magnetic field dominates the energy density in the emitting region. Thus, the dominant emission mechanism will be synchrotron emission from relativistic electrons, since no cooling mechanism is known which is faster \citep[see, e.g.,][]{Beniamini14a}. Our observational results therefore also pose a challenge to Poynting flux dominated models, although Compton up-scattering from seed photons in the environment of an emerging Baryon-free jet offer a potential means of combining strongly magnetic outflows with a thermalized component or sharp spectrum \citep[see][for a recent example]{Gill14a}. Moreover, \citet{Beloborodov13a} argues that other optically thin emission models share the same problems of the synchrotron emission models, e.g., pitch-angle synchrotron radiation \citep{Baring04a} when the scatter angle in the comoving frame is not isotropic, and jitter radiation in turbulent magnetic fields \citep{Medvedev00a}.

\begin{figure*}
\resizebox{\hsize}{!}
{\includegraphics{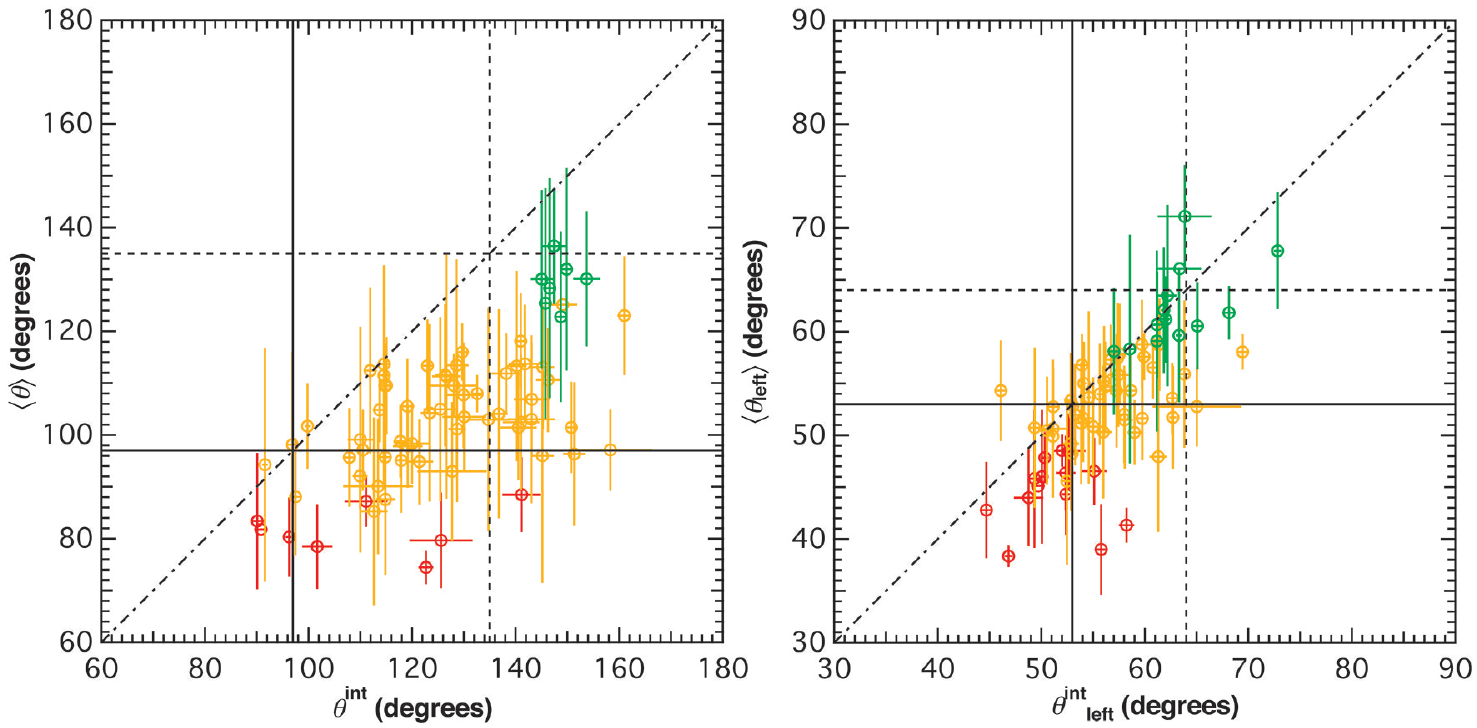}}
\caption{Left panel: Comparison between the average sharpness angles, $\langle \theta \rangle$, to the sharpness angles computed using the time-integrated catalog, $\theta^\text{int}$. Right panel: Comparison between the average left angles, $\langle \theta_\text{left} \rangle$, to the left angles computed using the time-integrated catalog, $\theta^\text{int}_\text{left}$. The dash-dotted line shows $x=y$. The solid and dashed lines show the single-electron synchrotron and Maxwellian synchrotron limit, respectively. We note that the error bars of $\langle \theta \rangle$ and $\langle \theta_\text{left} \rangle$ represent the spread in $\theta$ and $\theta_\text{left}$. See main text for the color-coding and details about the plots.}
\label{fig:average}
\end{figure*}

Finally, we compute the average time-resolved sharpness angles and left angles, $\langle \theta \rangle$ and $\langle \theta_\text{left} \rangle$, weighing each spectrum equally. In Fig.~\ref{fig:average}, we compare $\langle \theta \rangle$ and $\langle \theta_\text{left} \rangle$ to the sharpness angles and left angles computed using the time-integrated catalog \citep{Gruber14a}, $\theta^\text{int}$ and $\theta^\text{int}_\text{left}$, for every burst in our sample (listed also in Table~\ref{tab:average}). In the left panel, green color indicates the 7 bursts (10\%) whose average sharpness angles are consistent with the Maxwellian synchrotron limit (individual $\theta$ values can still be inconsistent, see, e.g., GRB 090829.672 from Fig.~\ref{fig:lcs}), orange color indicates the 55 bursts (79\%) that are inconsistent with the Maxwellian synchrotron limit but consistent on average with the single-electron synchrotron limit, and red color indicates the 8 bursts (11\%) that are inconsistent with the single-electron synchrotron limit. Similarly, in the right panel, green color indicates the 13 bursts (19\%) whose average sharpness angles are consistent with the Maxwellian synchrotron limit, orange color indicates the 43 bursts (61\%) that are inconsistent with the Maxwellian synchrotron limit but consistent on average with the single-electron synchrotron limit, and red color indicates the 14 bursts (20\%) that are inconsistent with the single-electron synchrotron limit. We note that the error bars of $\langle \theta \rangle$ and $\langle \theta_\text{left} \rangle$ represent the standard deviations $\textit{SD} = \sqrt{\langle \theta^2 \rangle - \langle \theta \rangle^2}$ and $\textit{SD}_\text{left} = \sqrt{\langle \theta^2_\text{left} \rangle - \langle \theta_\text{left} \rangle^2}$, which indicate the spread of the angle distributions within each burst. The error bars of $\theta^\text{int}$ and $\theta^\text{int}_\text{left}$ are computed using the same procedure as described in Sect.~\ref{sect:resu}, and are relatively small because the parameters are better constrained by higher photon counts.

Figure~\ref{fig:average} shows that the time-integrated angles are systematically larger than the average time-resolved angles for individual bursts, and the data points lie closer to (or sometimes even above) the diagonal in the right panel. One reason for this is that the spectral evolution of $E_\text{p}$ is corrected for when computing $\langle \theta \rangle$, but not when computing $\theta^\text{int}$. Another reason is that rotation of the triangle between spectra, where a decrease in $\theta_\text{left}$ is compensated for by an increase in  $\theta_\text{right} =  \theta - \theta_\text{left}$, or vice versa (i.e., a specific joint  change in power-law indices), is also corrected for when computing $\langle \theta \rangle$. This latter compensation is not possible for $\langle \theta_\text{left} \rangle$, and the the data points in the right panel of Fig.~\ref{fig:average} therefore lie closer to the diagonal. We also emphasise that different light curve binning methods are used in the time-resolved and time-integrated spectral catalog. In our time-resolved analysis, as mention in Sect.~\ref{sect:anal}, the light curves are binned with $S/N=30$, and then those spectra without a peak or break are excluded. In the time-integrated catalog \citep[see, e.g.,][]{Gruber14a}, all time intervals with $S/N \ge 3.5$ are included. The fact that fewer bursts in the time-integrated spectral analysis are inconsistent with the Maxwellian limit (44 bursts, 62\% for $\theta^\text{int}$, and 62 bursts, 89\% for $\theta^\text{int}_\text{left}$) underlines the importance of time-resolved analysis.

\section{Summary and conclusions}
\label{sect:conc}

We have computed the sharpness angles $\theta$ of the observed time-resolved spectra of \textit{Fermi} GRBs, and compared the values to the sharpest cases of the synchrotron radiation theory, namely the single-electron synchrotron and the Maxwellian distributed synchrotron emission function. We find that over 91\% of the observed spectra are sharper than the Maxwellian synchrotron model predicts, indicating that synchrotron radiation cannot be responsible for the peaks or breaks of GRB prompt emission spectra. No general evolutionary trend is observed for $\theta$ within bursts. Moreover, the Maxwellian synchrotron function can only contribute up to $58_{-18}^{+23}$\% of the peak flux. We conclude that the underlying prompt emission mechanism in GRBs must produce spectra sharper than a Maxwellian synchrotron function but broader than a blackbody.

It is still possible for synchrotron emission to dominate the spectrum away from the peak or break observed in the GBM energy range (e.g., at the LAT energy range). Also, a sub-dominant synchrotron component can allow for a continuous connection to the afterglow phase, where synchrotron emission is typically dominant \citep[see, e.g.,][for a recent review]{vanEerten15a}. The transition between prompt and afterglow is then marked by the disappearance of the non-synchrotron (likely thermal) component. There are other theoretical possibilities to explain GRB prompt emission, such as the collisional model of electron-positron pairs \citep[e.g.,][]{Beloborodov10a}. For recent reviews on GRB prompt emission mechanisms, see, e.g., \citet{Zhang14a} and \citet{Peer15a}.

A possibly similar inference can be made on the related phenomena of prompt optical emission showing a similar temporal profile as the gamma-ray emission \citep[e.g.,][]{Elliott14a,Greiner14a} or very early X-ray flares (e.g., \citealt{Peer06a}; see also \citealt{Hu14a} for a recent large \textit{Swift} sample study): if the prompt emission is not dominated by synchrotron emission, this is likely the case for this longer wavelength emission as well \citep[see, e.g.,][]{Starling12a,Peng14a}.

We demonstrated in this paper a method to quantify the shape of the observed GRB spectra that provides a clear tool for distinguishing between various standard emission functions. Ultimately, the question as to the viability of any particular emission model can only be fully resolved if complete spectral predictions for that model are tested directly against photon counts \citep[see, e.g.,][]{Burgess14a}. Our paper predicts that any model based on standard synchrotron theory without additional radiative mechanisms will systematically struggle to capture the spectral curvature of the prompt emission. This will manifest itself in relatively poor CSTAT values and trends in the fit residuals.

\begin{acknowledgements}

The authors wish to thank Andrei Beloborodov, Alexander van der Horst, Asaf Pe'er, Bin-Bin Zhang, and Bing Zhang for insightful discussions, and Alexander Kann for proofreading the manuscript. HFY and JG acknowledge support by the DFG cluster of excellence "Origin and Structure of the Universe" (www.universe-cluster.de). HJvE acknowledges support by the Alexander von Humboldt foundation. RS is partially supported by ISF, ISA and iCore grants. The GBM project is supported by the German Bundesministeriums f{\"u}r Wirtschaft und Technologie (BMWi) via the Deutsches Zentrum f{\"u}r Luft und Raumfahrt (DLR) under the contract numbers 50 QV 0301 and 50 OG 0502.

\end{acknowledgements}

\bibliographystyle{aa} 
\bibliography{mybib.bib} 

\clearpage

\begin{appendix}

\section{Fitting functions}
\label{app:fitfunc}

The Comptonized model (COMP) is a power-law model with a high-energy exponential cutoff:
\begin{equation}
\label{eqn:comp}
f_\text{COMP}(E) = A \left(\frac{E}{100\text{ keV}}\right)^\alpha \exp\left[-\frac{(\alpha+2)E}{E_\text{p}}\right] \, ,
\end{equation}
where $A$ is the normalization factor at 100~keV in units of ph~s$^{-1}$~cm$^{-2}$~keV$^{-1}$, $\alpha$ is the power-law index, and $E_\text{p}$ is the peak energy in the $\nu F_\nu$ space in units of keV.

The Band function (BAND) is a model which a low-energy cutoff power law and a high-energy power law joined together by a smooth transition. It is an empirical function proposed by \citet{Band93a}:
\begin{equation}
\label{eqn:band}
f_\text{BAND}(E) = A\left\{
\begin{array}{ll}
	\left(\frac{E}{100\text{ keV}}\right)^\alpha \exp\left[-\frac{(\alpha+2)E}{E_\text{p}}\right]: E<E_\text{c} \, , \\
	\left(\frac{E}{100\text{ keV}}\right)^\beta \exp\left(\beta-\alpha\right) \left(\frac{E_\text{c}}{100\text{ keV}}\right)^{\alpha-\beta}: E\geq E_\text{c} \, ,
\end{array}
\right.
\end{equation}
where
\begin{equation}
\label{eqn:Ec}
E_\text{c}=\left(\frac{\alpha-\beta}{\alpha+2}\right)E_\text{p} \, .
\end{equation}
In Eqns.~(\ref{eqn:band}) and (\ref{eqn:Ec}), $A$ is the normalization factor at 100~keV in units of ph~s$^{-1}$~cm$^{-2}$~keV$^{-1}$, $\alpha$ is the low-energy power-law index, $\beta$ is the high-energy power-law index, $E_\text{p}$ is the peak energy in the $\nu F_\nu$ space in units of keV, and $E_\text{c}$ is the characteristic energy where the low-energy power law with an exponential cutoff ends and the pure high-energy power law starts, in units of keV. We note that when $\beta \to - \infty$ the Band function reduces to the Comptonized model.

The smoothly broken power law (SBPL) is a model of two power laws joined by a smooth transition. It was first parameterized by \citet{Ryde99a} and then re-parameterized by \citet{Kaneko06a}:
\begin{equation}
\label{eqn:sbpl}
f_\text{SBPL}(E) = A \left(\frac{E}{100\text{ keV}}\right)^b 10^{(a-a_\text{piv})} \, ,
\end{equation}
where
\begin{equation}
\label{eqn:sbplpara}
\left\{
\begin{array}{ll}
a = m \Delta \ln\left(\frac{e^q+e^{-q}}{2}\right), a_\text{piv} = m \Delta \ln\left(\frac{e^{q_\text{piv}}+e^{-q_\text{piv}}}{2}\right) \, , \\
m = \frac{\beta-\alpha}{2} \, , b = \frac{\alpha+\beta}{2} \, , \\
q = \frac{\log(E/E_\text{b})}{2} \, , q_\text{piv} = \frac{\log(100\text{ keV}/E_\text{b})}{2} \, .
\end{array}
\right.
\end{equation}
In Eqns.~(\ref{eqn:sbpl}) and (\ref{eqn:sbplpara}), $A$ is the normalization factor at 100~keV in units of ph~s$^{-1}$~cm$^{-2}$~keV$^{-1}$, $\alpha$ and $\beta$ are the low- and high-energy power-law indices respectively, $E_\text{b}$ is the break energy in units of keV, and $\Delta$ is the break scale. Unlike the Band function, the break scale is not coupled to the power-law indices, so SBPL is a five-parameter-model if we let $\Delta$ free to vary. It is fixed at $\Delta=0.3$ in all the \textit{Fermi} GBM GRB catalogs and is therefore adopted in this paper.

The peak energy of SBPL in the $\nu F_\nu$ space can be found at
\begin{equation}
\label{eqn:sbplep}
E_\text{p} = 10^x E_\text{b} \, , x = \Delta \tanh^{-1} \left(\frac{\alpha+\beta+4}{\alpha-\beta}\right) \, .
\end{equation}
We note that Eqn.~(\ref{eqn:sbplep}) is only valid for $\alpha>-2$ and $\beta<-2$.

\clearpage

\onecolumn

\section{Comparison between the average time-resolved and time-integrated sharpness angles}
\label{app:angles}

\LTcapwidth=\textwidth
\begin{longtable}{cccccccccc}
\caption{Comparison between the average time-resolved and the time-integrated sharpness angles. Column (1) lists the GRB names using the \textit{Fermi} GBM trigger designation that is assigned for each new trigger detected. Column (2) lists the numbers of spectra used in averaging $\theta$ and $\theta_\text{left}$, $N$, for individual bursts. Columns (3) and (5) list the average time-resolved sharpness angles $\langle \theta \rangle$ and left angles $\langle \theta_\text{left} \rangle$, and Cols. (4) and (6) list their respective standard deviations $\textit{SD} = \sqrt{\langle \theta^2 \rangle - \langle \theta \rangle^2}$ and $\textit{SD}_\text{left} = \sqrt{\langle \theta^2_\text{left} \rangle - \langle \theta_\text{left} \rangle^2}$. Columns (7) and (9) list the time-integrated sharpness angles $\theta^\text{int}$ and left angles $\theta^\text{int}_\text{left}$, and Cols. (8) and (10) list their respective errors.} \label{tab:average} \\
\hline\hline
GRB name & $N$ & $\langle \theta \rangle$ & $\textit{SD}$ & $\langle \theta_\text{left} \rangle$ & $\textit{SD}_\text{left}$ & $\theta^\text{int}$ & $\sigma^\text{int}$ & $\theta^\text{int}_\text{left}$ & $\sigma^\text{int}_\text{left}$ \\
(1) & (2) & (3) & (4) & (5) & (6) & (7) & (8) & (9) & (10) \\
\hline
\endfirsthead
\multicolumn{10}{c}%
{{\bfseries \tablename\ \thetable{} -- continued from previous page}} \\
\hline\hline
GRB name & $N$ & $\langle \theta \rangle$ & $\textit{SD}$ & $\langle \theta_\text{left} \rangle$ & $\textit{SD}_\text{left}$ & $\theta^\text{int}$ & $\sigma^\text{int}$ & $\theta^\text{int}_\text{left}$ & $\sigma^\text{int}_\text{left}$ \\
(1) & (2) & (3) & (4) & (5) & (6) & (7) & (8) & (9) & (10) \\
\hline
\endhead
\hline \multicolumn{10}{r}{\textit{Continued on next page}} \\
\endfoot
\hline
\endlastfoot
080817.161 & 14 & 113.72 & 11.40 & 58.10 & 6.09 & 141.78 & 3.11 & 57.02 & 0.53 \\
080825.593 & 12 & 104.97 & 17.79 & 50.46 & 5.20 & 125.47 & 3.03 & 50.57 & 0.68 \\
080916.009 & 12 & 97.13 & 7.84 & 52.77 & 3.85 & 158.31 & 8.04 & 65.02 & 4.35 \\
081125.496 & 4 & 79.65 & 9.23 & 43.98 & 4.68 & 125.60 & 5.95 & 48.77 & 1.37 \\
081207.680 & 7 & 102.97 & 21.57 & 50.74 & 7.73 & 134.76 & 3.85 & 49.36 & 0.67 \\
081215.784 & 16 & 113.42 & 20.45 & 51.99 & 4.89 & 128.58 & 2.33 & 53.27 & 0.35 \\
081224.887 & 11 & 88.08 & 11.32 & 48.23 & 5.65 & 97.49 & 0.89 & 52.98 & 0.44 \\
090131.090 & 1 & 125.13 & - & 66.06 & - & 149.10 & 2.88 & 63.36 & 2.07 \\
090328.401 & 7 & 109.53 & 9.37 & 58.74 & 4.45 & 115.07 & 0.90 & 61.40 & 0.42 \\
090424.592 & 27 & 94.86 & 8.24 & 51.65 & 4.05 & 121.41 & 2.46 & 59.76 & 0.40 \\
090528.516 & 9 & 113.35 & 8.92 & 60.55 & 4.21 & 122.99 & 1.15 & 65.08 & 0.53 \\
090530.760 & 4 & 74.46 & 3.24 & 41.34 & 1.68 & 122.69 & 1.43 & 58.23 & 0.74 \\
090618.353 & 47 & 118.12 & 9.18 & 61.19 & 4.17 & 140.99 & 0.93 & 62.03 & 0.39 \\
090626.189 & 10 & 110.62 & 10.07 & 55.91 & 7.11 & 146.27 & 2.43 & 63.82 & 0.64 \\
090718.762 & 7 & 98.76 & 7.02 & 53.58 & 3.43 & 117.73 & 1.39 & 62.65 & 0.65 \\
090719.063 & 19 & 80.31 & 7.63 & 44.33 & 3.92 & 96.27 & 0.77 & 52.38 & 0.38 \\
090804.940 & 1 & 81.78 & - & 45.12 & - & 90.82 & 1.23 & 49.69 & 0.61 \\
090809.978 & 7 & 102.46 & 13.01 & 51.64 & 4.24 & 141.06 & 3.46 & 54.05 & 0.87 \\
090820.027 & 82 & 95.73 & 10.20 & 49.91 & 4.23 & 114.84 & 1.32 & 51.12 & 0.23 \\
090829.672 & 32 & 130.12 & 13.03 & 67.80 & 5.62 & 153.69 & 2.56 & 72.83 & 0.38 \\
090902.462 & 60 & 98.14 & 17.93 & 50.32 & 6.40 & 96.89 & 1.05 & 56.00 & 0.85 \\
090926.181 & 56 & 104.08 & 20.17 & 50.84 & 4.76 & 136.77 & 1.80 & 55.01 & 0.42 \\
091003.191 & 11 & 111.50 & 21.26 & 59.10 & 8.73 & 114.55 & 1.02 & 61.16 & 0.48 \\
091010.113 & 3 & 87.61 & 14.55 & 47.94 & 7.25 & 114.84 & 1.81 & 61.30 & 0.85 \\
091120.191 & 5 & 104.85 & 6.26 & 56.53 & 3.03 & 113.75 & 1.39 & 60.79 & 0.65 \\
091127.976 & 4 & 136.45 & 11.17 & 71.11 & 4.98 & 147.43 & 2.08 & 63.86 & 2.56 \\
091128.285 & 5 & 92.07 & 6.28 & 50.28 & 3.11 & 109.96 & 1.42 & 59.00 & 0.67 \\
100322.045 & 17 & 101.47 & 8.81 & 54.31 & 4.86 & 150.77 & 0.97 & 46.10 & 0.51 \\
100324.172 & 9 & 83.40 & 13.16 & 45.83 & 6.68 & 90.10 & 0.83 & 49.33 & 0.41 \\
100414.097 & 14 & 94.27 & 22.51 & 46.01 & 6.45 & 91.59 & 0.55 & 50.07 & 0.27 \\
100511.035 & 2 & 107.98 & 3.59 & 58.05 & 1.71 & 132.58 & 1.06 & 69.44 & 0.48 \\
100612.726 & 4 & 98.33 & 7.75 & 46.38 & 3.58 & 120.01 & 3.33 & 52.36 & 0.96 \\
100707.032 & 21 & 96.00 & 24.53 & 39.00 & 4.37 & 145.17 & 2.17 & 55.78 & 0.52 \\
100719.989 & 13 & 111.32 & 23.61 & 45.57 & 8.05 & 126.63 & 2.91 & 52.48 & 0.53 \\
100724.029 & 40 & 128.33 & 21.23 & 53.66 & 8.31 & 146.58 & 1.07 & 54.59 & 0.22 \\
100728.095 & 20 & 87.19 & 4.95 & 47.85 & 2.48 & 111.12 & 4.14 & 50.33 & 0.39 \\
100826.957 & 14 & 131.98 & 19.54 & 53.98 & 4.89 & 149.83 & 0.91 & 55.65 & 0.30 \\
100829.876 & 4 & 88.51 & 7.22 & 48.49 & 3.63 & 141.16 & 3.82 & 52.67 & 1.64 \\
100910.818 & 5 & 103.49 & 10.35 & 55.83 & 5.01 & 130.08 & 4.46 & 57.44 & 1.01 \\
100918.863 & 26 & 101.71 & 8.23 & 55.00 & 4.02 & 99.82 & 0.58 & 54.12 & 0.28 \\
101014.175 & 37 & 122.80 & 16.43 & 62.07 & 6.07 & 148.72 & 1.09 & 61.83 & 0.28 \\
101023.951 & 10 & 111.86 & 7.84 & 58.76 & 4.30 & 138.19 & 2.83 & 59.73 & 0.86 \\
101123.952 & 30 & 113.28 & 18.37 & 55.25 & 5.31 & 140.21 & 1.95 & 56.06 & 0.29 \\
101126.198 & 11 & 116.00 & 5.50 & 61.82 & 2.58 & 129.69 & 1.25 & 68.13 & 0.57 \\
101231.067 & 3 & 92.98 & 13.39 & 50.64 & 6.66 & 127.74 & 6.73 & 51.13 & 1.44 \\
110301.214 & 21 & 105.55 & 9.15 & 54.71 & 4.33 & 119.12 & 1.28 & 56.21 & 0.47 \\
110407.998 & 4 & 95.66 & 9.52 & 52.03 & 4.68 & 107.89 & 1.19 & 58.01 & 0.57 \\
110428.388 & 5 & 78.48 & 8.18 & 42.80 & 4.64 & 101.69 & 3.00 & 44.70 & 0.54 \\
110622.158 & 9 & 104.27 & 17.11 & 54.30 & 3.55 & 123.40 & 1.64 & 58.65 & 0.54 \\
110625.881 & 37 & 101.15 & 14.11 & 51.18 & 5.87 & 128.70 & 1.67 & 53.85 & 0.44 \\
110717.319 & 12 & 106.86 & 11.01 & 57.45 & 5.26 & 143.11 & 3.61 & 57.38 & 0.69 \\
110721.200 & 9 & 123.02 & 11.45 & 58.32 & 11.02 & 160.99 & 1.31 & 58.56 & 0.51 \\
110729.142 & 6 & 113.66 & 10.91 & 60.67 & 5.08 & 114.58 & 1.15 & 61.18 & 0.54 \\
110731.465 & 7 & 90.09 & 13.06 & 49.21 & 6.50 & 113.45 & 6.95 & 52.88 & 0.77 \\
110817.191 & 3 & 96.32 & 13.85 & 46.54 & 3.21 & 151.33 & 2.06 & 55.12 & 1.14 \\
110825.102 & 18 & 95.10 & 10.12 & 51.74 & 4.95 & 117.89 & 0.81 & 62.72 & 0.38 \\
110920.546 & 11 & 85.26 & 18.09 & 38.35 & 1.06 & 112.64 & 2.66 & 46.85 & 0.38 \\
110921.912 & 9 & 109.51 & 6.17 & 57.30 & 3.38 & 128.10 & 4.55 & 56.72 & 0.49 \\
111003.465 & 5 & 101.43 & 10.22 & 51.49 & 4.69 & 140.55 & 3.51 & 58.05 & 0.84 \\
111216.389 & 7 & 112.41 & 15.99 & 57.59 & 2.28 & 111.91 & 1.23 & 59.92 & 0.58 \\
111220.486 & 9 & 109.50 & 8.12 & 58.74 & 3.86 & 114.75 & 0.82 & 61.26 & 0.38 \\
120119.170 & 14 & 107.72 & 10.16 & 57.87 & 4.81 & 129.94 & 3.60 & 57.56 & 0.67 \\
120129.580 & 26 & 97.85 & 8.05 & 52.85 & 4.05 & 119.18 & 2.69 & 53.05 & 0.41 \\
120204.054 & 39 & 111.53 & 13.74 & 59.63 & 6.46 & 126.38 & 2.02 & 63.29 & 0.31 \\
120226.871 & 9 & 113.08 & 11.90 & 56.78 & 2.98 & 145.26 & 2.20 & 53.93 & 0.56 \\
120328.268 & 22 & 125.40 & 22.21 & 53.35 & 4.62 & 145.76 & 1.12 & 52.85 & 0.43 \\
120426.090 & 10 & 99.11 & 21.79 & 48.54 & 1.58 & 110.01 & 2.42 & 52.02 & 0.67 \\
120526.303 & 10 & 97.10 & 7.83 & 52.75 & 3.80 & 110.52 & 3.43 & 51.15 & 0.48 \\
120624.933 & 18 & 103.02 & 16.19 & 54.30 & 5.48 & 143.06 & 4.30 & 57.26 & 1.26 \\
120707.800 & 12 & 130.08 & 17.19 & 63.49 & 8.75 & 145.04 & 2.13 & 62.19 & 0.93 \\

\hline
\end{longtable}
\twocolumn

\end{appendix}

\end{document}